\documentclass[amsmath,amssymb,aps,pra,preprint,groupedaddress]{revtex4-1}

\usepackage[dvipdfmx]{graphicx}
\usepackage{dcolumn}
\usepackage{bm}
\usepackage{longtable}
\usepackage{ulem}

\usepackage{xcolor,soul}
\usepackage{mathptmx}
\usepackage{color,amsmath}%
\usepackage{ccaption}

\newcommand*\Laplace{\mathop{}\!\mathbin\bigtriangleup}

\begin{document}

\title{Numerical study of anisotropic diffusion in Turing patterns based on Finsler geometry modeling}

\author{Gildas Diguet}
\affiliation{%
Micro System Integration Center, Tohoku University, Sendai, Japan
}%
\author{Madoka Nakayama }
\affiliation{%
Research Center of Mathematics for Social Creativity, Research Institute for Electronic Science, Hokkaido University, Sapporo, Japan
}%
\author{Sohei Tasaki}
\affiliation{%
Department of Mathematics, Faculty of Science, Hokkaido University, Sapporo, Japan
}%

\author{Fumitake Kato}%
\author{Hiroshi Koibuchi} 
\email[Corresponding author: ]{koibuchi@gm.ibaraki-ct.ac.jp; koi-hiro@sendai-nct.ac.jp}
\affiliation{
National Institute of Technology (KOSEN), Ibaraki College, Hitachinaka, Japan.
}
\date{\today} 

\author{Tetsuya Uchimoto}
\affiliation{%
Institute of Fluid Science (IFS), Tohoku University, Sendai, Japan
}%
\affiliation{ELyTMaX, CNRS-Universite de Lyon-Tohoku University, Sendai, Japan}

\begin{abstract}
We numerically study the anisotropic Turing patterns (TPs) of an activator-inhibitor system, focusing on anisotropic diffusion by using the Finsler geometry (FG) modeling technique. In the FG modeling prescription, the diffusion coefficients are dynamically generated to be direction dependent owing to an internal degree of freedom (IDOF) and its interaction with the activator and inhibitor under the presence of thermal fluctuations.
In this sense, FG modeling contrasts sharply with the standard numerical technique, where direction-dependent diffusion coefficients are assumed in the reaction-diffusion (RD) equations of Turing. To find the solution of the RD equations, we use a hybrid numerical technique as a combination of the metropolis Monte Carlo method for IDOF updates and discrete RD equations for steady-state configurations of activator-inhibitor variables.
We find that the newly introduced IDOF and its interaction are one possible origin of spontaneously emergent anisotropic patterns on living organisms such as zebra and fishes. Moreover, the IDOF makes TPs controllable by external conditions if the IDOF is identified with lipids on cells or cell mobility.

\end{abstract}

\maketitle

\section{Introduction\label{intro}}
Turing patterns (TPs) are described by partial differential equations of Turing \cite{Turing-TRSL1952} using two different variables $u$ and $v$, which are scalar functions on a domain in ${\bf R}^2$ in the case of a two-dimensional system. These $u$ and $v$ parameters are usually called the activator and inhibitor, respectively \cite{Koch-Meinhardt-RMP1994,FitzHugh-BP1961,Nagumo-etal-ProcIRE1962}, owing to their interaction properties as implemented in the reaction and diffusion (RD) terms in the equation. TPs emerge on scales ranging from macroscopic \cite{,Sekimura-etal-PRSL2000,Kondo-Miura-Science2010,Bullara-Decker-NatCom2015} to microscopic \cite{Tan-etal-Science2018,Fuseya-etal-NatPhys2021}.

These patterns emerge as a result of competition between diffusion and reaction, and therefore, many studies have been conducted to extend the Laplace operators of the RD equation to the graph Laplacian to find TPs in random networks \cite{Nakao-etal-NatPhys2010,Carletti-Nakao-PRE2020,Asllani-etal-NatCom2014,Asllani-etal-PRE2014,Petit-etal-PhysA2016}. In those networks, topological properties rather than geometric properties play a significant role in forming TPs. Pattern formation depends on the node number, which is the total number of connections, and TPs can be visualized by using a ``coordinate axis'' of node numbers. Another extension is to modify diffusivity to accommodate non-Gaussian behavior of Brownian particles confined in narrow plates by including fluctuations in diffusion constants, where position-dependent and anisotropic diffusion constants are assumed \cite{Alexandre-etal-PRL2023}.
Such a non-Gaussian distribution of particle displacements is considered anomalous diffusion corresponding to anomalous transport phenomena in crowded biological materials such as cellular membranes \cite{Bressloff-Newby-RMP2013,Hofling-Franosch-RPP2013,Sokolov-etal-PhysToday2002,Metzler-Klafter-JPhysA2004}. This anomalous transport is characterized by subdiffusion, which is described by a power law behavior of the mean-square displacement $\sim t^\alpha (0\!<\!\alpha\!<\!1)$, observed at intermediate time scales. So-called super diffusion characterized by $ t^\alpha (\alpha\!>\!1)$ is observed in bacterial swarming \cite{Ariel-etal-NJP2013,Ariel-etal-NatCom2015}. This phenomenon is described by the Levy walk, which is a model of a random walk with a constant speed.

\begin{figure}[h!]
\centering{}\includegraphics[width=12.5cm]{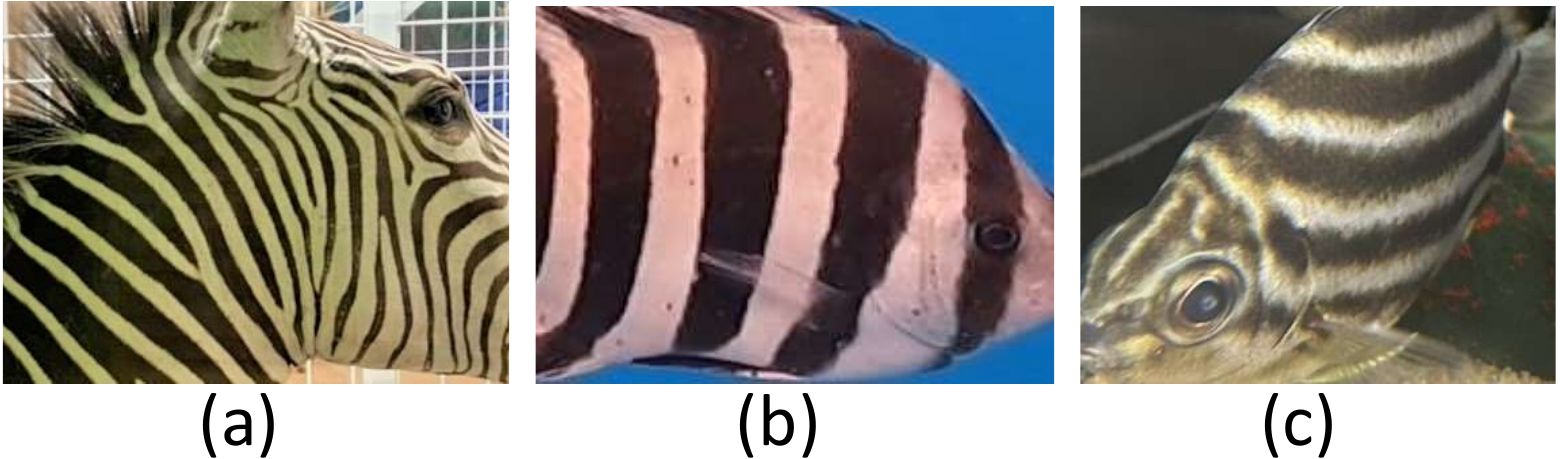}
\caption{ Anisotropic Turing patterns on (a) zebra and (b), (c) fishes. The patterns vary depending on the individual, while the direction of patterns depends only on the species. \label{fig-1}  }
\end{figure}
Anisotropic TPs observed on zebra and fishes (Fig. \ref{fig-1}) are known to emerge as a consequence of a difference in diffusion constants between the activator and inhibitor. This difference in diffusion constants corresponding to marine angelfish was estimated by Kondo and Asai in Ref. \cite{Kondo-Nature1995}. In Refs. \cite{Shoji-etal-DevDyn2003,Iwamoto-Shoji-RIMS2018}, Shoji et al. reported that anisotropy in diffusion constants is effective in determining the direction of stripe patterns in numerical studies. TPs appear on curved surfaces \cite{Varea-etal-PRE1999}.
Krause et al. reported that anisotropy in the evolution of TPs is sensitive to curvature on growing domains, which are two-dimensional curved surfaces embedded in ${\bf R}^3$, assuming the induced metric for describing the Laplace Beltrami operator in the diffusion terms \cite{Krause-etal-BulMatBiol2019}.

Anisotropic diffusion is ubiquitously expected in many phenomena. Particles in cubic crystals under a stress field undergo direction-dependent diffusion \cite{Dederichs-Schroeder-PRB1978}, and anisotropic surface diffusion of CO on Ni(110) has been experimentally observed \cite{Xiao-etal-PRL1991}. Light propagates faster in one direction than in the other directions in strongly scattering media \cite{Bret-Lagendijk-PRE2004}, and
skyrmion transport is also direction dependent, accompanying a shape deformation due to an applied in-plane magnetic field \cite{Kerber-etal-PRAP2021}. All these phenomena, including anisotropic TPs such as those in Fig. \ref{fig-1}, are well described by assuming direction-dependent diffusion constants because the underlying origins of anisotropy are mostly well understood. However, such a phenomenological description is not always satisfactory and can be further improved.

In this paper, we study the origin of anisotropic diffusion in TPs. For this purpose, we focus on a mechanism of dynamic anisotropy in which diffusion constants dynamically appear as direction-dependent constants, allowing us to understand anisotropic diffusion without assuming these constants \cite{ICMsquare2020-JPCconf2022,Koibuchi-etal-ICMsquare2022}. Here, we recall that such dynamic anisotropy is successfully implemented in interaction coefficients in several statistical mechanical models using the Finsler geometry (FG) modeling technique \cite{Koibuchi-Sekino-PhysA2014,Takano-Koibuchi-PRE2017,Proutorov-etal-JPC2018,ElHog-etal-PRB2021,ElHog-etal-RinP2022}. This technique simply changes length scales for interactions from the Euclidean length to position- and direction-dependent lengths along the local coordinate systems, and detailed information on new interactions is unnecessary. 
Thus, we consider that the FG modeling technique takes advantage of phenomenological modeling. Indeed, FG modeling prescription is applicable to Laplace operators in the diffusion terms of the RD equation. This applicability comes from the fact that differential equation models such as RD equations share the same property in their interactions with those statistical mechanical models in which FG modeling is successful. In addition, the stochastic nature in the internal degree of freedom (IDOF) introduced in Monte Carlo (MC) methods is shared with the convergent configurations of activator-inhibitor variables, and hence, physical quantities are obtained by calculating the mean values of many convergent steady-state configurations of the RD equation.

This paper is organized as follows: In Section \ref{Std-approach}, we review the RD equation of the FN type for the variables $u$ and $v$ from a numerical point of view and show numerical data such as snapshots of isotropic and anisotropic TPs obtained on a regular square lattice of size $N\!=\!100^2$, where anisotropic TPs appear as a result of assumed direction-dependent diffusion coefficients in the RD equations. To quantify the anisotropic TPs and evaluate the effect of the direction-dependent diffusion coefficients, we introduce absolute second-order derivatives of $u$ and $v$ as well as squares of the first-order derivatives. In Section \ref{FG-modeling}, we introduce the FG modeling technique to implement the dynamical anisotropy in the Laplace operator in the RD equation by including a new IDOF on two types of triangulated lattices: fixed-connectivity and dynamically triangulated lattices. A numerical technique, which we call the hybrid technique, is introduced to update $u$ and $v$ and the new IDOFs. In Section \ref{results}, we show numerical data including snapshots obtained on these two lattices. Finally, we summarize the results in Section \ref{conclusion}, where remaining problems are also mentioned.

\section{Standard approach to Turing patterns \label{Std-approach}}
\begin{figure}[h!]
\centering{}\includegraphics[width=12.5cm]{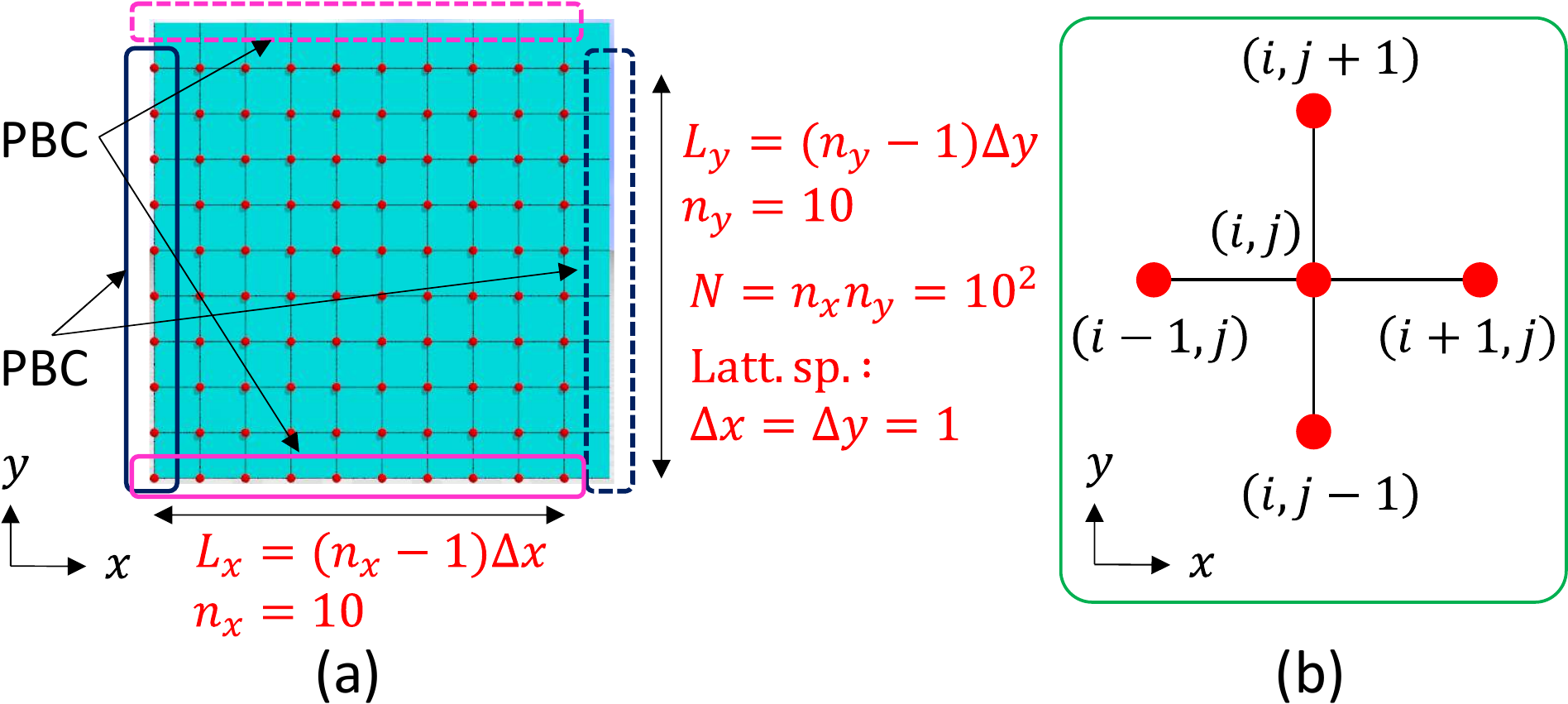}
\caption { (a) Regular square lattice with PBCs of size $N\!=\!10^2$, which is small compared with $N\!=\!100^2$ assumed in the simulations in the following subsection, and (b) lattice site $(i,j)$ and its four nearest neighbor sites, where $1\!\leq\!i\!\leq\!n_x, 1\!\leq\!j\!\leq\!n_y$. The lattice spacing is assumed to be $\Delta x\!=\!\Delta y\!=\!1$ in the simulations.
\label{fig-2} }
\end{figure}

In this section, we review TPs described by the FitzHugh-Nagumo (FN) equation and show snapshots of isotropic and anisotropic TPs and related physical quantities, which we call absolute second-order partial differentials, obtained by standard numerical techniques on regular square lattices with periodic boundary conditions (PBCs).
In Fig. \ref{fig-2}(a), we show a small lattice of size $N\!=\!10^2$ to visualize the lattice structure, and a lattice site $(i,j)$ and its four neighboring sites are illustrated in Fig. \ref{fig-2}(b). Only a regular square lattice is considered in this section. 
\subsection{FitzHugh-Nagumo equation with diffusion anisotropy}
Let $u(x,y)$ and $v(x,y)$ be the variables corresponding to the activator and inhibitor, respectively, which satisfy the RD equations of FN type
\begin{eqnarray}
\begin{split}
\label{FN-eq-Eucl}
&\frac {\partial u}{\partial t}=D_u {\Laplace}u +f(u ,v), \quad  f=u -u ^3-v, \\
&\frac {\partial v}{\partial t}=D_v  \Laplace v+\gamma g(u,v), \quad g= u -\alpha v
\end{split}
\end{eqnarray}
on the two-dimensional plane \cite{Shoji-etal-DevDyn2003,Iwamoto-Shoji-RIMS2018}. The first and second terms on the right-hand side are called the diffusion and reaction terms, respectively, where $\Laplace\!=\!\frac{\partial^2}{\partial x^2}\!+\!\frac{\partial^2}{\partial x^2}$ is the Laplace operator. The symbols $D_u$ and $D_v$ are the diffusion coefficients, and $\alpha$ and $\gamma$ are constants.

For suitable ranges of the parameters, the RD equations have certain steady-state solutions called TPs with a periodicity in spatial directions. When the periodicity appears almost regularly in one direction, the patterns become anisotropic, as shown in Fig. \ref{fig-1}(a)--(c). These anisotropic patterns can be reproduced by using the RD equation with diffusion anisotropy introduced with the parameters $a$ and $b$ such that \cite{Iwamoto-Shoji-RIMS2018}
\begin{eqnarray}
\begin{split}
\label{modified-Laplacian}
&\Laplace u \to {\Laplace}_a u=a\frac{\partial^2 u }{\partial x^2}+(2-a)\frac{\partial^2u}{\partial y^2},\; (0<a<2), \\
&\Laplace v \to {\Laplace}_b v=b\frac{\partial^2 v }{\partial x^2}+(2-b)\frac{\partial^2v}{\partial y^2},\; (0<b<2). 
\end{split}
\end{eqnarray}
For the ranges $0\!<\!a\!<\!2$ and $0\!<\!b\!<\!2$, the diffusion constants $D_u$ and $D_v$ in Eq. (\ref{FN-eq-Eucl}) effectively becomes direction dependent such that
\begin{eqnarray}
\label{direction-dep-diff-constants}
\begin{split}
&D_u \to\left (D_u^x,D_u^y\right)=\left(aD_u, (2-a)D_u\right), \\
&D_v \to \left(D_v^x,D_v^y\right)=\left(bD_u, (2-b)D_v\right). 
\end{split}
\end{eqnarray}
The direction-dependent coefficients can be represented by the ratio
\begin{eqnarray}
\label{direction-dep-coefficients}
c^{u}_{v}=\frac{c_u}{c_v}=\frac{a(2-b)}{b(2-a)}, \quad {\rm where} \quad  c_u=\frac{D_u^x}{D_u^y}=\frac{a}{2-a}, \quad c_v=\frac{D_v^x}{D_v^y}=\frac{b}{2-b}.
\end{eqnarray}
We call $c^{u}_{v}$ the anisotropy coefficient or simply anisotropy.
Note that the conditions $a\!=\!b\!=\!1$ correspond to the isotropic diffusion represented by the anisotropy $c^{u}_v\!=\!1$, which implies isotropy. The condition $c^{u}_v\!=\!1$ is expected for all $a\!=\!b$ even when $a\!\not=\!1$, where $c_u\!\not=\! 1$ and $c_v\!\not=\! 1$.

\subsection{Numerical solutions on a regular square lattice with periodic boundary conditions }
The time evolution equations in Eq. (\ref{FN-eq-Eucl}) are replaced by the discrete time evolution equations using the discrete time step $\Delta t$ such that
\begin{eqnarray}
\label{discrete-t-iterations}
\begin{split}
&u_{ij}(t+{\Delta} t)\leftarrow u_{ij}(t) +{\Delta} t \left[D_u{\Laplace}_au_{ij}(t) +f\left(u_{ij}(t),v_{ij}(t)\right)\right],\\
&v_{ij}(t+{\Delta} t)\leftarrow v_{ij}(t) +{\Delta} t \left[D_v{\Laplace}_bv_{ij}(t) +g\left(u_{ij}(t),v_{ij}(t)\right)\right], 
\end{split}
\end{eqnarray}
where $u_{ij}$ and $v_{ij}$ denote the discrete analogs of $u(x,y)$ and $v(x,y)$ defined at lattice site $(i,j)$, $(1\!\leq\!i\!\leq\!n_x, 1\!\leq\!j\!\leq\!n_y)$ (Fig. \ref{fig-2}(a)). The discrete diffusion terms ${\Laplace}_au_{ij}$ and ${\Laplace}_bv_{ij}$ are given by
\begin{eqnarray}
\label{discrete-Laplace-on-sq-latt}
\begin{split}
& {\Laplace}_au_{ij}=\frac{a}{(\Delta x)^2}\left(u_{i+1,j}+u_{i-1,j}-2u_{i,j}\right)+\frac{2-a}{(\Delta y)^2}\left(u_{i,j+1}+u_{i,j-1}-2u_{i,j}\right),\\
& {\Laplace}_bv_{ij}= \frac{b}{(\Delta x)^2}\left(v_{i+1,j}+v_{i-1,j}-2v_{i,j}\right)+\frac{2-b}{(\Delta y)^2}\left(v_{i,j+1}+v_{i,j-1}-2v_{i,j}\right)
\end{split}
\end{eqnarray}
for all $(i,j)$ with lattice spacing $\Delta x$ and $\Delta y$, where the positions $(i\!\pm\! 1,j)$ and $(i,j\!\pm\! 1)$ are shown in Fig. \ref{fig-2}(b). The convergent criteria of the iterations in Eqs. (\ref{discrete-t-iterations}) are given by
\begin{eqnarray}
\label{convergence-std}
\begin{split}
&{\rm Max}\left\{|u_{ij}(t+\Delta t)-u_{ij}(t)|\right\} < 1\times 10^{-8}, \\
&{\rm Max}\left\{|v_{ij}(t+\Delta t)-v_{ij}(t)|\right\} < 1\times 10^{-8}. 
\end{split}
\end{eqnarray}

\begin{figure}[h!]
\centering{}\includegraphics[width=13.5cm]{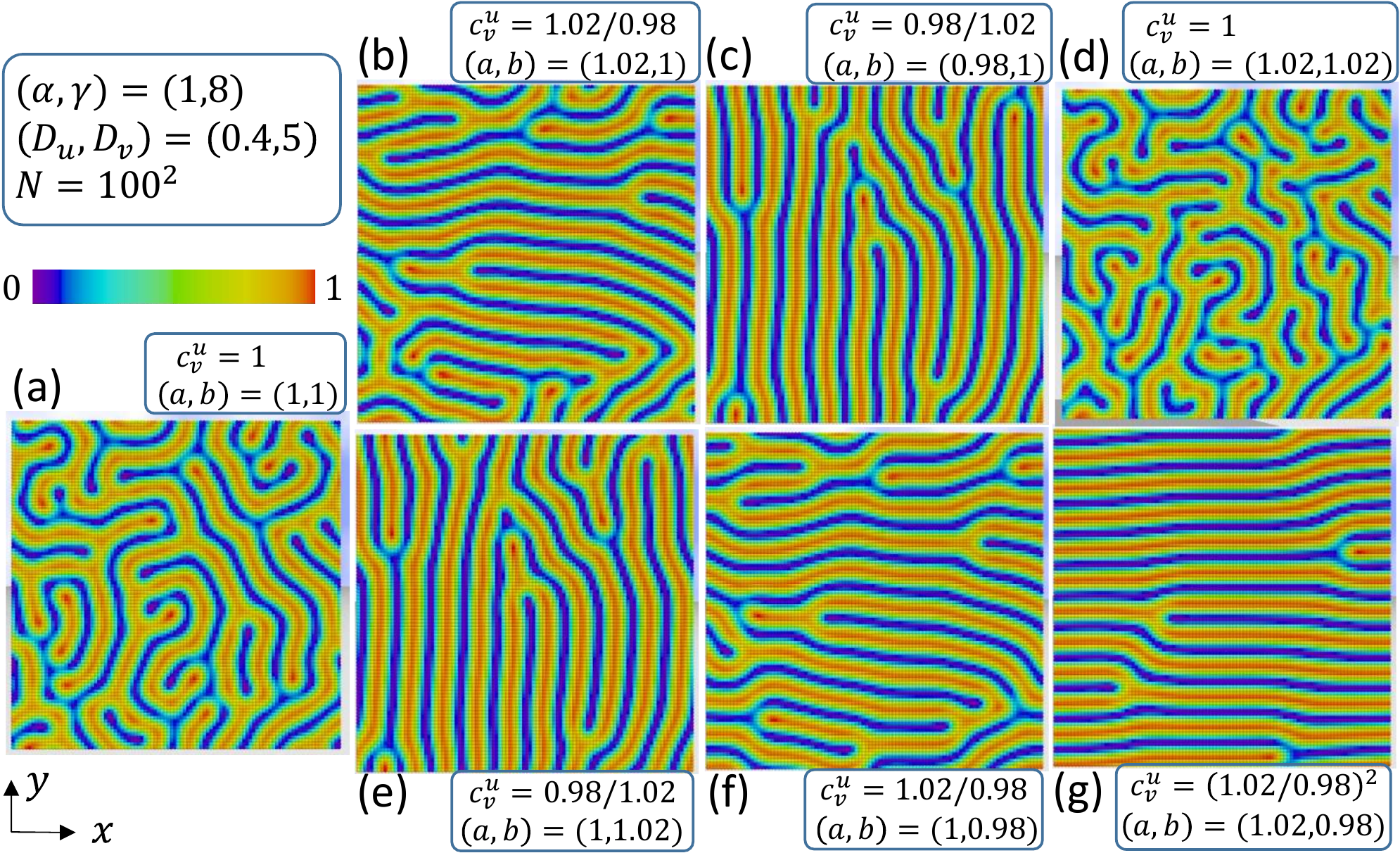}
\caption {Snapshots of the standard model on a regular square lattice of size $N\!=\!100^2$ with PBCs. The anisotropy parameters $(a,b)$ fixed to (a) $(1,1)$, (b) $(1.02,1)$, (c) $(0.98,1)$, (d) $(1.02,1.02)$, (e) $(1,1.02)$, (f) $(1,0.98)$ and (g) $(1.02,0.98)$, and the anisotropies $c_v^u$ in Eq. (\ref{direction-dep-coefficients}) are shown. The pattern is isotropic when $a\!=\!b$, as in (a) and (d), represented by $c_v^u\!=\!1$, while the stripe lies along the $x$ direction when $a\!>\!b$, as in (b), (f) and (g), represented by $c_v^u\!>\!1$, and it lies along the $y$ direction when $a\!<\!b$, as in (c) and (e), represented by $c_v^u\!<\!1$.
\label{fig-3} }
\end{figure}
Snapshots of the convergent configurations of $u$ on the lattice of size $N\!=\!100^2$ ($\Leftrightarrow n_x\!=\!n_y\!=\!100$) are plotted in Fig. \ref{fig-3}(a)--(g), where $\Delta t\!=\!1\times 10^{-3}$ and $\Delta x\!=\!\Delta y\!=\!1$.
The value of $u$ in the snapshots is normalized to $[0,1]$ to visualize the patterns, and the corresponding color code is shown, as are the assumed parameters. We confirm that the patterns are isotropic when $a\!=\!b$ in (a) $a\!=\!b\!=\!1$ and (d) $a\!=\!b\!=\!1.02$, while they are anisotropic when $a\!>\!b$ in (b) and (f) and $a\!<\!b$ in (c) and (e). This change from isotropy to anisotropy is enhanced when the anisotropy $c_v^u$ is increased from $c_v^u\!=\!1$ in (a) to $c_v^u\!=\!1.02/0.98$ in (b) and $c_v^u\!=\!(1.02/0.98)^2$ in (g). We find from (b) and (f) that the increase ($\nearrow$) in $a$ and the decrease ($\searrow$) in $b$ cause the same effect, enforcing the pattern anisotropy along the $x$ direction, and from (c) and (e) that the decrease ($\searrow$) of $a$ and the increase ($\nearrow$) of $b$ cause the same effect, enforcing the pattern along the $y$ direction. These effects are consistent with the observation that the anisotropy of the pattern in (g) is stronger than that in (b) and (f), where $c_v^u\!=\!(1.02/0.98)^2$ in (g) is larger than $c_v^u\!=\!1.02/0.98$ in (b) and $c_v^u\!=\!1.02/0.98$ in (f). Snapshots of $v$ are almost identical to those of $u$ and are not plotted.

To quantify the observed anisotropy in the TPs caused by anisotropic diffusion constants in the diffusion terms of Eq. (\ref{discrete-Laplace-on-sq-latt}), we calculate the mean values of the absolute of the second-order partial differentials of $u$ and $v$
\begin{eqnarray}
\label{absolute-partial2-on-sq-latt}
\begin{split}
&d^2_x u=\frac{1}{N}\sum_{ij}\left|(u_{i+1,j}+u_{i-1,j}-2u_{i,j}\right|, \quad d^2_y u=\frac{1}{N}\sum_{ij}\left|u_{i,j+1}+u_{i,j-1}-2u_{i,j}\right|,\\
&d^2_x v=\frac{1}{N}\sum_{ij}\left|v_{i+1,j}+v_{i-1,j}-2v_{i,j}\right|, \quad d^2_y v=\frac{1}{N}\sum_{ij}\left|v_{i,j+1}+v_{i,j-1}-2v_{i,j}\right|,
\end{split}
\end{eqnarray}
which are ``interaction'' parts, while the factors $a$, $2\!-\!a$ and $b$, $2\!-\!b$ in Eq. (\ref{discrete-Laplace-on-sq-latt}) are considered ``coefficients'', from which $D_u$ and $D_v$ are removed for simplicity. These interactions and coefficients are closely connected to each other. Therefore, the interactions are expected to be dependent on the parameters $a$ and $b$ and, as a consequence, on the input anisotropy $c_v^u$.
The reason the absolute symbol $|*|$ is used is that the sign of the second-order partial differentials $\partial^2_{x,y}u$ and $\partial^2_{x,y}v$ are locally changeable in their sign and cancel out when they are summed over, and hence, we need $|*|$ in the sum to see how much $\partial^2_{x,y}u$ and $\partial^2_{x,y}v$ deviate from zero.

\begin{figure}[h!]
\centering{}\includegraphics[width=11.5cm]{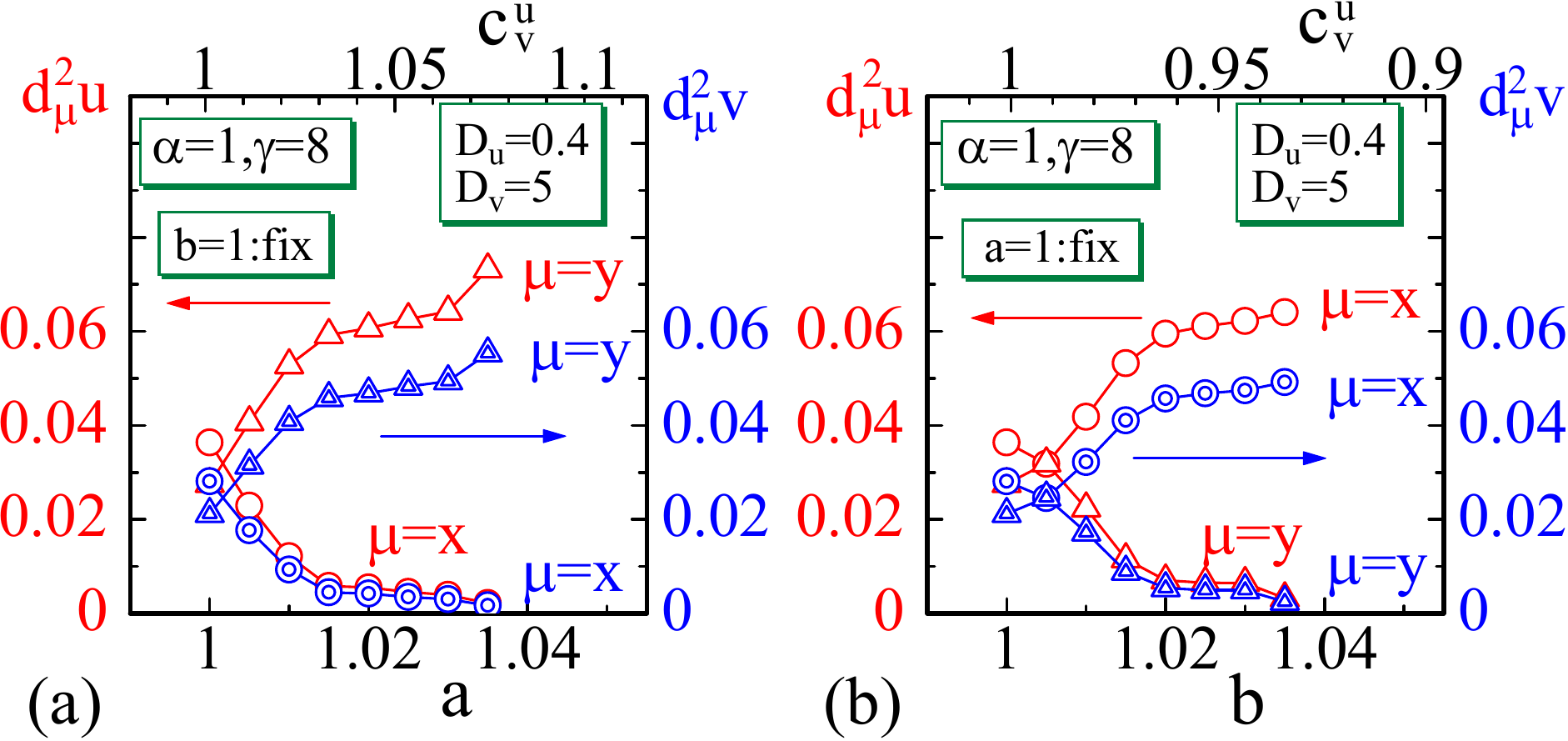}
\caption {Absolute second-order partial differentials $d^2_\mu u$ and $d^2_\mu v$ obtained with (a) $a$ variation under the conditions $a\!\geq\!b(\!=\!1)$ and (b) $b$ variation under $(1\!=\!)a\!\leq\!b$ on the $N\!=\!100^2$ lattice. The horizontal lower axes represent $a$ and $b$. Axes $a$ and $b$ are more intuitive than $c_v^u$, which is also shown on the upper axes.
The results $d^2_x u\!<\!d^2_y u$ and $d^2_x v\!<\!d^2_y v$ obtained under the conditions $a\!\geq\!b(=\!1)$ in (a) and those $d^2_x u\!>\!d^2_y u$ and $d^2_x v\!>\!d^2_y v$ obtained under $(1\!=\!)a\!\leq\!b$ in (b) are consistent with the pattern anisotropy in the $x$ and $y$ directions in Fig. \ref{fig-3}, respectively, as discussed in the text.
\label{fig-4} }
\end{figure}
Figure \ref{fig-4}(a)  shows $d^2_\mu u$ and $d^2_\mu v$ ($\mu\!=\!x,y$) calculated by varying $a$ under the conditions $b\!=\!1$ and $a\!\geq\! 1$. The other input parameters shown in the figure are the same as those assumed in the snapshots in Fig. \ref{fig-3}. We find from Fig. \ref{fig-4}(a) that $d^2_xu$ and $d^2_xv$ decrease ($\searrow$) and that $d^2_yu$ and $d^2_yv$ increase ($\nearrow$) with increasing $a$ ($\nearrow$). The results plotted in Fig. \ref{fig-4}(b) also show how $d^2_\mu u$ and $d^2_\mu v$ vary when $b$ increases ($\nearrow$) under the conditions $a\!=\!1$ and $b\!\geq\! 1$.
These results are summarized as follows:
\begin{eqnarray}
\label{results-std-1}
\begin{split}
{\rm Fig. \ref{fig-4}(a)(b=1):}\; &d_x^2u \searrow\; ({\rm if}\; a\nearrow(>1)),\;\;d_y^2u \nearrow  \; ({\rm if}\; 2-a\searrow(<1)),\\
\left[\Leftrightarrow \;\right.&\left.d_x^2u \nearrow\; ({\rm if}\; a\searrow(<1)),\;\;d_y^2u \searrow  \; ({\rm if}\; 2-a\nearrow(>1)),\right]\\
&d_x^2v \searrow\; (b=1),\;\;d_y^2v \nearrow  \; (2-b=1),
\end{split}
\end{eqnarray}
and
\begin{eqnarray}
\label{results-std-2}
\begin{split}
{\rm Fig. \ref{fig-4}(b)(a=1):}\; &d_x^2u \nearrow\; (a=1),\;\;d_y^2u \searrow  \; (2-a=1),\\
&d_x^2v \nearrow\; ({\rm if}\; b\nearrow(>1)),\;\;d_y^2v \searrow  \; ({\rm if}\; 2-b\searrow(<1)).\\
\left[\Leftrightarrow \;\right.&\left.d_x^2v \searrow\; ({\rm if}\; b\searrow(<1)),\;\;d_y^2v \nearrow  \; ({\rm if}\; 2-b\nearrow(>1)).\right]
\end{split}
\end{eqnarray}
The statements in the second line of Eq. (\ref{results-std-1}) and those in the third line of Eq. (\ref{results-std-2}) have no corresponding data in Fig. \ref{fig-4}(a) and (b) and appear not to be supported by the obtained data. However, this outcome is expected because anisotropic patterns depend only on the ratio $a/b$ at least in the region close to $a/b\!=\!1$ under the condition $a\!=\!1$ or $b\!=\!1$. For this reason, we remove the data for $a\!<\!1, b\!=\!1$ and $b\!<\!1, a\!=\!1$ from Fig. \ref{fig-4}(a) and (b) to simplify the discussion.

The behaviors of $d^2_\mu u$ and $d^2_\mu v$ plotted in Fig. \ref{fig-4}(a) and (b) are expected from the snapshots in Fig. \ref{fig-3}. Indeed, we expect $d^2_xu \!<\! d^2_yu$ in the patterns that are anisotropic along the $x$ direction, such as those in Fig. \ref{fig-3}(b), (f) and (g). On the patterns showing anisotropy along the $x$ direction, $u$ is almost constant, implying a small $d^2_xu$.
This behavior $d^2_xu \!<\! d^2_yu$ is also expected from the fact that the direction-dependent diffusion constants $\left (D_u^x,D_u^y\right)\!=\!\left(aD_u, (2-a)D_u\right)$ in Eq. (\ref{direction-dep-diff-constants}) are $D_u^x>D_u^y$ for $a\!>\!1$ and $b\!=\!1$ because a large diffusion constant $D_u^x$ induces a long-distance spatial correlation of $u$, implying almost constant $u$ along the $x$-axis, which is reflected in a small $d^2_xu$. In fact, if the long-range correlation of $u$ is accompanied by a large $d^2_xu$, which corresponds to a large value of the integral of $(\partial_x u)^2$ over the domain, then the corresponding energy $\int dxdy ((\partial_x u)^2\!+\!(\partial_y u)^2)$ becomes very large. This contradicts the fact that Turing instability lies close to a stable state, which is the energy minimum, of solutions in the RD equation.

On the other hand, $d^2_{x,y}v$ has a notably similar behavior to $d^2_{x,y}u$ even under the isotropic constants $D_v^x\!=\!D_v^y\!=\!D_v$ due to the condition $b\!=\!1$, as shown in the third line of Eq. (\ref{results-std-1}). This nontrivial behavior in $d^2_{x,y}v$ comes from the implemented interaction between $u$ and $v$ in the reaction terms $f$ and $g$ in Eq. (\ref{FN-eq-Eucl}). We note that this interaction is present without the diffusion anisotropy $(a,b)\!=\!(1,1)$, and therefore, this interaction is a reason for the deviations $d^2_xu \!\not=\!d^2_yu$ at $a\!=\!1$ and $d^2_xv \!\not=\!d^2_yv$ at $b\!=\!1$ in Fig. \ref{fig-4}(a) and (b). These small deviations imply that the patterns are not completely isotropic but slightly anisotropic and spontaneously generated, even though the snapshot in Fig. \ref{fig-3}(a) looks isotropic.

The result obtained under the conditions $a\!>\! 1$ and $b\!=\!1$ indicates that the patterns of $u$ anisotropy in the $x$ direction are caused by the assumed diffusion anisotropy, and moreover, the anisotropic patterns in $v$ are caused by both the anisotropic pattern of $u$ and the interaction between $u$ and $v$.
The data $d^2_\mu u$ and $d^2_\mu v$ plotted in Fig. \ref{fig-4}(b) are calculated by varying $b$ with fixed $a\!=\!1$. In this case, the condition $a\!<\! b$ is satisfied and opposite to $a\!>\! b$ in Fig. \ref{fig-4}(a), and hence, we obtain the results $d^2_xu \!>\! d^2_yu$ and $d^2_xv \!>\! d^2_yv$, which are opposite to those obtained in Fig. \ref{fig-4}(a), as summarized in Eq. (\ref{results-std-2}). The result confirms that the anisotropic patterns in $u$ are caused by the anisotropic patterns of $v$ and the interaction between $u$ and $v$.

\begin{figure}[h!]
\centering{}\includegraphics[width=11.5cm]{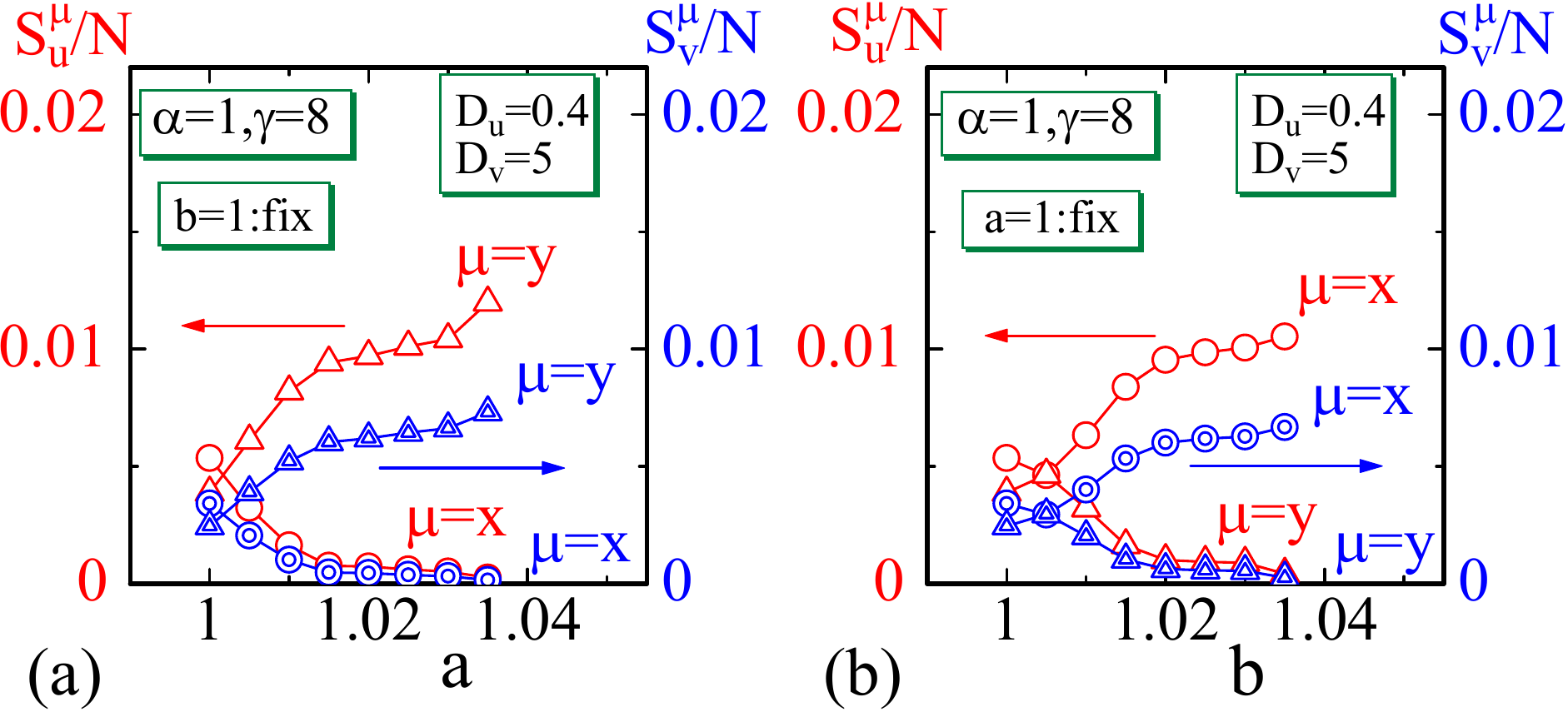}
\caption {$S^\mu_u, (\mu=x,y)$ and  $S^\mu_v, (\mu=x,y)$ obtained with (a) $a$ variation under the conditions $a\!\geq\!b(\!=\!1)$ and (b) $b$ variation under  $(1\!=\!)a\!\leq\!b$ on the $N\!=\!100^2$ lattice. The assumed parameters are the same as those in Fig. \ref{fig-4}.
\label{fig-5}}
\end{figure}
Anisotropies in diffusion constants in Eq. (\ref{direction-dep-diff-constants}) can also be reflected in the quantities $\int (\frac{\partial u}{\partial \mu})^2dxdy, (\mu\!=\!x,y)$ and $\int (\frac{\partial v}{\partial \mu})^2dxdy, (\mu\!=\!x,y)$. The discrete expressions, denoted by $S^\mu_u$ and $S^\mu_v$, are given by
\begin{eqnarray}
\label{direc-dep-energies}
\begin{split}
&S^x_u=\frac{1}{4}\sum_{ij} \left(u_{i+1,j}-u_{i-1,j}\right)^2, \quad S^y_u=\frac{1}{4}\sum_{ij} \left(u_{i,j+1}-u_{i,j-1}\right)^2,\\
&S^x_v=\frac{1}{4}\sum_{ij} \left(v_{i+1,j}-v_{i-1,j}\right)^2, \quad S^y_v=\frac{1}{4}\sum_{ij} \left(v_{i,j+1}-v_{i,j-1}\right)^2,
\end{split}
\end{eqnarray}
where differentials are replaced by differences; $\frac{\partial u}{\partial x} \to \frac{1}{2\Delta x}\left(u_{i+1,j}-u_{i-1,j}\right)$ with $\Delta x\!=\!1$.
The behaviors of $S^\mu_u/N$ and $S^\mu_v/N$ plotted in Fig. \ref{fig-5}(a) and (b) are almost the same as those in Fig. \ref{fig-4}(a) and (b).

\section{Finsler geometry modeling of Turing patterns\label{FG-modeling}}
In the preceding section, we confirm that anisotropic TPs appear when the diffusion constants are anisotropic in the DR equations under the condition of suitable input parameters. However, as emphasized in the introduction, the origin of such anisotropy in the diffusion constants is unclear.
In this section, which is the main part of this paper, we show one possible origin of the anisotropic diffusion constants.

\subsection{Internal degree of freedom on the triangulated lattice and the Monte Carlo update\label{App-A}}
\begin{figure}[h]
\centering{}\includegraphics[width=12.5cm,clip]{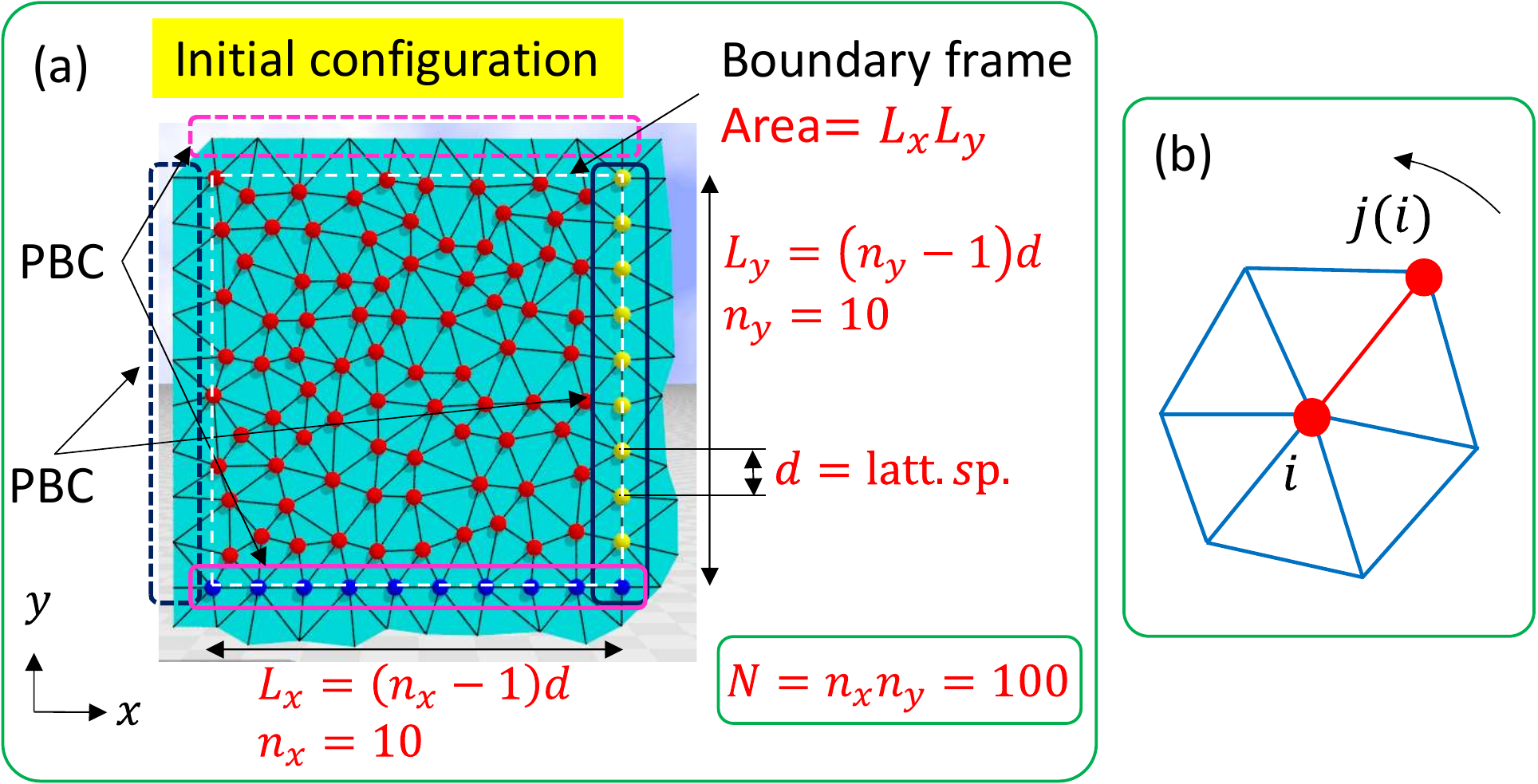}
\caption{
(a) Triangulated square lattice of size $N\!=\!100$, which is the total number of vertices given by $N\!=\!n_xn_y$, where the total number of boundary vertices is $n_\mu\!=\!10(\mu\!=\!x,y)$. The edge length $L_\mu (\mu\!=\!x,y)$ is given by $L_\mu\!=\!(n_\mu\!-\!1)d$, where $d$ is the lattice spacing. PBCs are assumed on the boundary vertices. The dashed square is the boundary frame of area $A\!=\!L_xL_y$ implemented by the PBCs. (b) A lattice site $i$ and its neighboring sites denoted by $j(i)$, $1\!\leq\!j(i)\!\leq\!J_{\rm max}$ with $J_{\rm max}\!=\!6$. This $J_{\rm max}$ depends on the site $i$ and is also called the coordination number denoted by $q_i$.
\label{fig-6}}
\end{figure}
First, we present detailed information on triangulated lattices, in which the vertex position plays a role in the IDOF for anisotropic diffusion. This new IDOF is governed by the Hamiltonian $S$, as introduced in the following subsection.
A triangulated lattice is plotted in Fig. \ref{fig-6}, where PBCs are assumed for all vertex positions because their positions are allowed to move as shown below. This assumption is slightly different from those associated the regular square lattice in Fig. \ref{fig-2}(a), where PBCs are assumed only on the boundary vertices.
On the initial configuration in Fig. \ref{fig-6}(a), vertices enclosed by the solid rectangles are identified with those enclosed by the same-sized oblong dashed rectangles, which are plotted at the positions $n_\mu d, (\mu\!=\!x,y)$ distant from the original positions, where $n_\mu$ and $d$ stand for the total number of vertices on the lattice edges and the lattice spacing suitably assumed in the simulations, respectively. The dashed square represents an implemented frame by PBCs, in which the vertex position $\vec{r}$ is identified with $\vec{r}\!\pm\!(n_xd,n_yd)$:
\begin{eqnarray}
\label{PBC}
\vec{r} \equiv \vec{r}\pm(n_xd,n_yd)\quad ({\rm PBCs}),
\end{eqnarray}
Consequently, the frame area is fixed to $A\!=\!L_xL_y$, where $L_\mu (\mu\!=\!x,y)$ is the edge length given by $L_\mu\!=\!(n_\mu\!-\!1)d$. A frame tension or surface tension thus emerges on the surface, as discussed in the following section.
Vertices on and inside the dashed frame in Fig. \ref{fig-6}(a) represent the initial configuration, and those inside are randomly distributed under the constraint that the minimum distance $r_{\rm min}\!=\!0.8d$ from the other vertices under the PBCs in Eq. (\ref{PBC}). This $r_{\rm min}\!=\!0.8d$ is assumed only in the initial lattice construction. The vertices are linked with the Voronoi tessellation technique \cite{Friedberg-Ren-NPB1984}.

%
\begin{figure}[h]
\centering{}\includegraphics[width=12.5cm,clip]{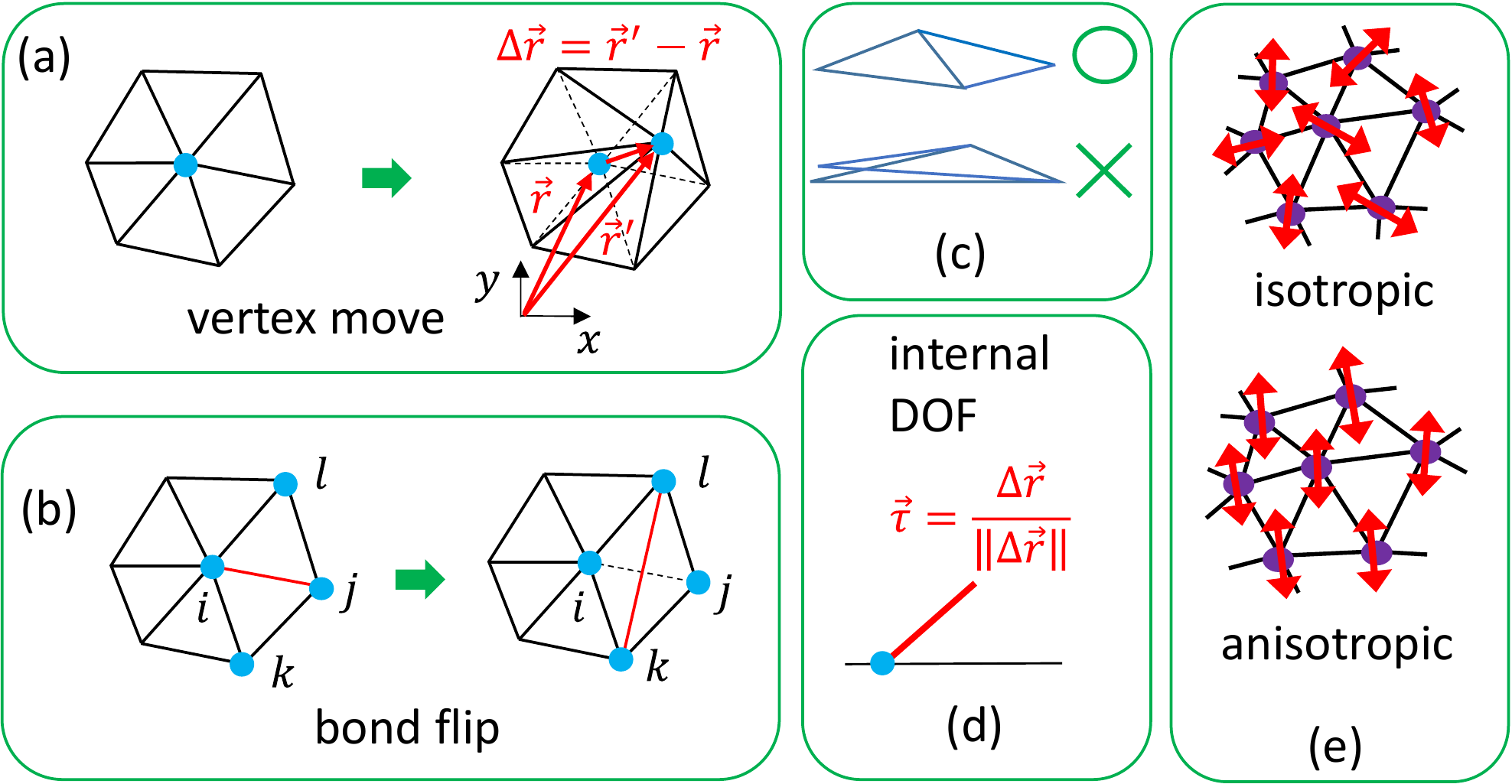}
\caption{
Illustrations of MC updates for the (a) vertex position and (b) bond flip under the constraint of nonfolding triangles. Lattices in which only the vertex position is updated are called fixed connectivity lattices, and those in which both the vertex position and the bond flip are updated are called fluid lattices. (c) Triangles are allowed to deform in no-folding configurations ($\bigcirc$) and prohibited from folding ($\times$). (d) The internal degree of freedom $\vec{\tau}(\in S^1/2)$ at $\vec{r}$ is defined by $\vec{\tau}\!=\!{\Delta} \vec{r}/\|{\Delta} \vec{r}\|$ using the deformation vector ${\Delta} \vec{r}\!=\!\vec{r}^\prime\!-\!\vec{r}$. (e) Illustrations of vertex fluctuations corresponding to isotropic and anisotropic configurations of $\vec{\tau}$.
\label{fig-7} }
\end{figure}
Next, to introduce an IDOF, we briefly explain the MC update \cite{Metropolis-JCP-1953,Landau-PRB1976} of the vertex position $\vec{r} (\in {\bf R}^2)$, which is illustrated in Fig. \ref{fig-7}(a). The new position $\vec{r}^{\,\prime}$ is accepted with probability ${\rm Min}[1,\exp(\delta S)]$, where $\delta S\!=\!S(\vec{r}^{\,\prime})\!-\!S(\vec{r})$ is the energy change before and after the vertex movement $\vec{r}\!\to\! \vec{r}^{\,\prime}$. The new position $\vec{r}^{\,\prime}$ is randomly fixed in a small circle of radius $R$ centered at $\vec{r}$, where the radius $R$ is fixed so that the acceptance rate is approximately equal to $60\%-90\%$ under the constraint that the triangles are not folded at any bonds (nonfolding and folding triangles are illustrated in Fig. \ref{fig-7}(c)). This MC process for the update of vertex positions is the same as in MC studies for polymerized membranes \cite{KANTOR-NELSON-PRA1987,Gompper-Kroll-PRA1992,KOIB-PRE-2005}, which is a two-dimensional extension of the linear chain model for polymers \cite{Doi-Edwards-1986}.
Another MC process is the so-called bond flip procedure in Fig. \ref{fig-7}(b) \cite{Ho-Baum-EPL1990}. In this process, the bond $ij$ connecting vertices $i$ and $j$ is removed, and vertices $k$ and $l$ are connected by a bond. This bond flip is also accepted with probability ${\rm Min}[1,\exp(\delta S)]$ with energy change $\delta S$ under the constraint for the nonfolding triangles. The IDOF $\vec{\tau}$ is defined by $\vec{\tau}\!=\!{\Delta} \vec{r}/\|{\Delta} \vec{r}\| (\in S^1/2)$ (Fig. \ref{fig-7}(d)), which has values on the half circle $S^1/2$ due to the nonpolar nature assumed on the variable. Figure \ref{fig-7}(e) illustrates isotropic and anisotropic configurations of  $\vec{\tau}$.

\begin{figure}[h]
\centering{}\includegraphics[width=9.5cm,clip]{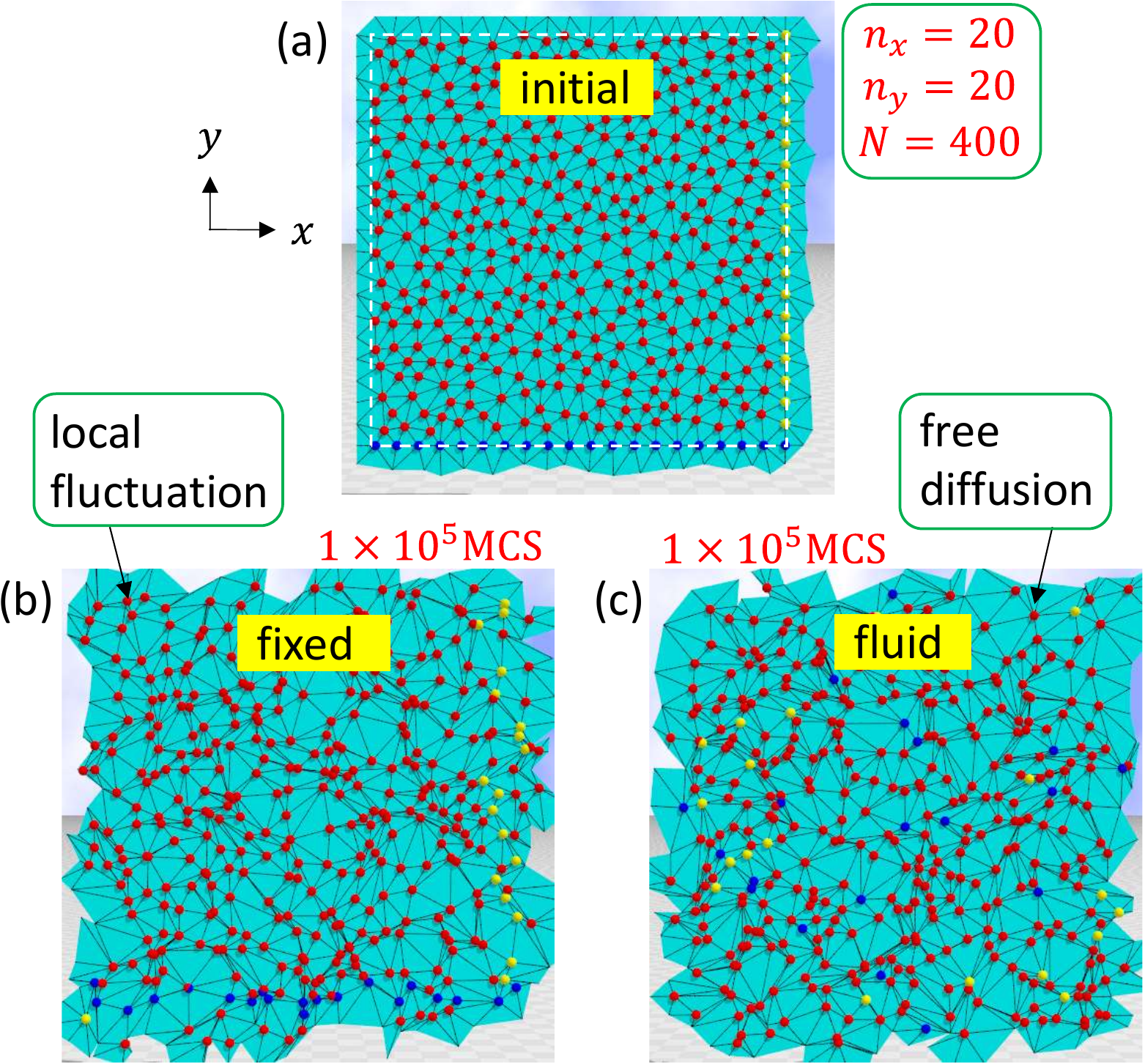}
\caption{
Snapshots of vertex configurations updated by the metropolis MC procedure of lattice size $N\!=\!400$: (a) initial configuration without the MC update, (b) fixed connectivity lattice obtained after $1\!\times\!10^5$ MCS for the vertex updates, and (c) fluid lattice obtained after $1\!\times\!10^5$ MCS for the updates of vertex and bond flip. Vertices on the boundaries of the fixed lattice in (b) fluctuate only locally, while those of the fluid lattice in (c) diffuse over the lattice. The total number of bonds $N_B$ and the total number of triangles $N_T$ are given by $N_B\!=\!3N$ and $N_T\!=\!2N$ on triangulated lattices with PBCs satisfying zero Euler characteristics $N\!-\!N_B\!+\!N_T\!=\!0$. \label{fig-8} }
\end{figure}
Finally, in this subsection, we show snapshots of vertex configurations updated by MC processes, including the initial configuration (Fig. \ref{fig-8}(a)). To clearly show the vertices, we use a lattice of size $N\!=\!400$. The snapshot denoted by ``fixed'' in Fig. \ref{fig-8}(b) is obtained after $1\!\times\!10^5$ MC sweeps (MCSs) for the update of vertex position $\vec{r}_i$ only, where 1 MCS represents $N$ consecutive updates of $\vec{r}_i, (i\!=\!1,\cdots,N)$. Vertices outside the dashed square, which is shown only in Fig. \ref{fig-8}(a), are plotted at the opposite side due to the PBC. We find that the vertices fluctuate only locally on the fixed connectivity lattice. On the other hand, the vertices diffuse almost freely over the lattice denoted by ``fluid'' in Fig. \ref{fig-8}(c) \cite{Ho-Baum-EPL1990,Peliti-Leibler-PRL1985}; we find that the boundary vertices are randomly mixed on the fluid lattice, where 1 MCS represents $N$ consecutive updates of vertex positions followed by $N$ random updates of bond flips.

We emphasize that these two different types of triangulated lattices are not simply different in the polygon shape from the regular square lattice in Section \ref{Std-approach} but rather have extra dynamic degrees of freedom, such as the vertex position. To suitably treat this new IDOF, we introduce the so-called Gaussian bond potential, usually assumed for a model of cell membranes \cite{HELFRICH-1973,Peliti-Leibler-PRL1985,NELSON-SMMS2004}, in the following subsection. Note also that IDOF $\vec{\tau}$ is connected to directions of the position movements and hence is called ``internal'', in contrast to the variables $u(x,y)$ and $v(x,y)$, which are functions of the positions.

\subsection{Finsler geometry modeling of anisotropic diffusion}
Now, we introduce the FG modeling technique for anisotropic TPs based on the DR equations of the FN type in Eq. (\ref{FN-eq-Eucl}). FG modeling modifies length scales for interactions to be direction dependent by using an IDOF that is not included in the original system, and therefore, the FG model is an extension of the original model owing to the new IDOF. As a consequence, the interaction coefficients in the original system are dynamically direction dependent. The original system in this paper is the DR equations in Eq. (\ref{FN-eq-Eucl}), which contain the variables $u$ and $v$. The time evolution of these variables is also iterated in the FG modeling by using the DR equations for obtaining TPs, while the new IDOF $\vec{\tau}$ is updated by the MC procedure because we have no differential equation for $\vec{\tau}$. To perform the MC procedure for $\vec{\tau}$, we assume the following discrete Hamiltonian $S$ composed of a linear combination of several terms such that
\begin{eqnarray}
\label{discrete-total-Hamiltonian}
\begin{split}
&S(\vec{r};\vec{\tau})=\left\{ \begin{array}{@{\,}ll}
                 S_1+U_V+D_u S_{u}+D_v S_{v}+\lambda S_\tau+S_\tau^F, \quad  ({\rm fixed}) \\
                 S_1+U_V+U_{\cal T}+D_u S_{u}+D_v S_{v}+\lambda S_\tau+S_\tau^F, \quad  ({\rm fluid})
                  \end{array} 
                   \right., 
\end{split}
\end{eqnarray}
where $S(\vec{r};\vec{\tau})$ denotes that the variables $\vec{r}$ and $\vec{\tau}$ are not independent; $\vec{\tau}$ is defined by the fictitious time evolution of $\vec{r}$ in the MC update as shown in Fig. \ref{fig-7}(c). The difference in $S$ between the fixed and fluid models in Eq. (\ref{discrete-total-Hamiltonian}) is that a potential $U_{\cal T}$ is included in $S$ for the fluid model, where the symbol ${\cal T}$ denotes a triangulation, which is considered as a variable only in the fluid model (Fig. \ref{fig-7}(b)). The partition functions are written as
\begin{eqnarray}
\label{part-funct}
\begin{split}
 &Z_{\rm fix}=\int \prod_i d\vec{r}_i \exp(-S), \quad (\textrm {fixed model}), \\
 &Z_{\rm flu}=\sum_{\mathcal T}\int \prod_i d\vec{r}_i \exp(-S), \quad (\textrm {fluid model}),
 \end{split}
\end{eqnarray}
where $\prod_i d\vec{r}_i$ denotes $2N$-dimensional multiple integrations on a domain $\cal{D}$ in ${\bf R}^2$ and where $\sum_{\mathcal T}$ in $Z_{\rm flu}$ denotes the sum over all possible triangulations. These are simulated by the MC updates in Fig. \ref{fig-7}(a) and (b).

The terms on the right-hand side of $S$ are defined as follows:
\begin{eqnarray}
\label{discrete-Hamiltonian}
\begin{split}
&S_1=\sum_{ij}\ell_{ij}^2,\quad \ell_{ij}^2\!=\!(\vec{r}_i\!-\!\vec{r}_j)^2,\\
&U_V=\sum_{ij}U_{ij}, \quad U_{ij}=\left\{ \begin{array}{@{\,}ll}
                 0\quad  (\ell_{\min}\leq \ell_{ij}\leq \ell_{\rm max}) \\
                 \infty\quad  ({\rm otherwise})
                  \end{array} 
                   \right., \\
&\quad(\ell_{\min}=0.01d  \quad\ell_{\rm max}=3d,\quad d : {\rm latt.\; sp.}),\\
&U_{\cal T}=\sum_{i}U_{i}, \quad U_{i}=\left\{ \begin{array}{@{\,}ll}
                 0\quad  (4\leq q_{i}\leq 9) \\
                 \infty\quad  ({\rm otherwise})
                  \end{array} 
                   \right., \\
&S_{u}=\sum_{ij}\gamma_{ij}^u\left(u _j-u _i\right)^2, \quad \gamma_{ij}^u=\frac{1}{6}\left(\frac{\chi_{ij}^u}{\chi_{ik}^u}+\frac{\chi_{ji}^u}{\chi_{jk}^u}+\frac{\chi_{ij}^u}{\chi_{il}^u}+\frac{\chi_{ji}^u}{\chi_{jl}^u}\right),\\
&S_{v}=\sum_{ij}\gamma_{ij}^v\left(v _j-v _i\right)^2, \quad \gamma_{ij}^v=\frac{1}{6}\left(\frac{\chi_{ij}^v}{\chi_{ik}^v}+\frac{\chi_{ji}^v}{\chi_{jk}^v}+\frac{\chi_{ij}^v}{\chi_{il}^v}+\frac{\chi_{ji}^v}{\chi_{jl}^v}\right),\\
&  S_\tau=-\sum_{ij}\left(\vec{\tau}_i\cdot \vec{\tau}_j\right)^2, \quad S_\tau^F=-\sum_i\left(\vec{\tau}_i\cdot {\vec F}\right)^2, \quad {\vec F}=(F_x,F_y)
\end{split}
\end{eqnarray}
The first term $S_1$ is the Gaussian bond potential, which is a spring potential defined by the sum of bond length squares $\ell_{ij}^2$, and $\sum_{ij}$ denotes the sum over bonds $ij$. The meaning of the inclusion of $S_1$ in $S$ is that the domain $\cal{D}$, defined by a fixed frame of side length $L_{x}$ and $L_{y}$ for PBCs in Fig. \ref{fig-6}(a), can be regarded as a membrane surface with an internal structure rather than a plane in ${\bf R}^2$ \cite{KANTOR-NELSON-PRA1987,Gompper-Kroll-PRA1992,KOIB-PRE-2005,Ho-Baum-EPL1990,HELFRICH-1973,Peliti-Leibler-PRL1985,NELSON-SMMS2004}.
Therefore, we consider that the frame is spanned by a membrane of area $A\!=\!L_xL_y$ with surface tension $\sigma$, as mentioned in the preceding subsection (see Appendix \ref{App-A} for detailed information on $\sigma$).

The second term $U_V$ is a constraint potential that prohibits the bond length $\ell_{ij}$ from being out of the range $\ell_{\min}\leq \ell_{ij}\leq \ell_{\rm max}$ with $\ell_{\min}\!=\!0.01d$ and $\ell_{\rm max}\!=\!3d$. The lattice spacing $d$ is fixed to two different values in the simulations, as discussed in Section \ref{results}, to check the influence of the surface tension $\sigma$ (Appendix \ref{App-A}).
The vertex moves are constrained inside $\cal{D}$ by the fixed frame of side lengths $L_{x}$ and $L_{y}$ in Fig. \ref{fig-6}(a) and by the condition for nonfolding triangles in Fig. \ref{fig-7}(c). These constraints can also be written in constraint potentials; however, we omit these potentials for simplicity.
We note that $S_1/N\!=\!1$ is satisfied in the case without any constraints for the bond length, such as the constraint for the side lengths $L_{x}$ and $L_{y}$ (Fig. \ref{fig-6}(a)) and the constraint $U_V$, due to the scale-invariant property of the partition function (see Appendix \ref{App-A} for further detail).

The term $U_{\cal T}$ for the fluid model is a constraint potential that enforces $4\leq q_{i}\leq 9$, where the coordination number $q_i$ is the total number of bonds connected to vertex $i$ (Fig. \ref{fig-6}(b)). The terms $S_{u}$ and $S_{v}$ are discrete versions of the continuous Hamiltonians
\begin{eqnarray}
\label{cont-Hamiltonian}
S_{u}= \frac{1}{2}\int \sqrt{g^u}d^2x \,g^{u,ab}\frac{\partial u }{\partial x^a}\frac{\partial u }{\partial x^b}, \quad S_{v}= \frac{1}{2} \int \sqrt{g^v}d^2x \,g^{v,ab}\frac{\partial v }{\partial x^a}\frac{\partial v }{\partial x^b}. 
\end{eqnarray}
In these expressions, $g^{u,ab}$ and $g^{v,ab}$ are the inverse and $g^{u,v}$ is the determinant of the Finsler metrics (see Appendix \ref{App-B} for further detail on the Finsler metric)
\begin{eqnarray}
\label{Finsler-metric}
g^u_{ab}= 
\begin{pmatrix}
1/(\chi_{ij}^u)^{2} & 0 \\ 
0 & 1/(\chi_{ik}^u)^{2} 
\end{pmatrix}, 
  \quad
 g^v_{ab}= 
\begin{pmatrix}
1/(\chi_{ij}^v)^{2} & 0  \\
0 & 1/(\chi_{ik}^v)^{2} 
\end{pmatrix},
\end{eqnarray}
where $\chi_{ij}^u$ and $\chi_{ij}^v$ are given by
\begin{eqnarray}
\label{Finsler-unit-lengths}
\begin{split}
&\chi_{ij}^u=|\vec{\tau}_i\cdot\vec{e}_{ij}| + \chi_0,\\
&\chi_{ij}^v=\sqrt{1-|\vec{\tau}_i\cdot\vec{e}_{ij}|^2} + \chi_0.
\end{split}
\end{eqnarray}
These continuous Hamiltonians in Eq. (\ref{cont-Hamiltonian}) correspond to the diffusion terms in Eq. (\ref{FN-eq-Eucl}). The position- and direction-dependent coefficients $\gamma_{ij}^u$ and $\gamma_{ij}^v$ in Eq. (\ref{discrete-Hamiltonian}) are given in Appendix \ref{App-B}. $S_\tau$ represents the interaction energy for nearest neighbor pairs of $\vec{\tau}_i$ and $\vec{\tau}_j$, which are assumed to be nonpolar, with the interaction coefficient $\lambda$. One additional assumption is that some external force $\vec{F}$ aligns $\vec{\tau}$ along the direction of $\vec{F}$. This interaction is described by the final term $S_\tau^F$, where nonpolar interaction is assumed. If the interaction is connected to electromagnetic interactions, polar interactions can also be assumed in $S_\tau$ and $S_\tau^F$.

Here, we comment on an implication implemented in the definitions of $\chi_{ij}^u$ and $\chi_{ij}^v$ in Eq. (\ref{Finsler-unit-lengths}). Let us suppose $\vec{e}_{ij}$ is parallel to the $x^1$-axis. Then, the definition of $\chi_{ij}^u$ implies that the Finsler length $ds_u\!=\!dx^1/\chi_{ij}^u$ (Appendix \ref{App-B}) for interactions in $u$ becomes short along the direction of $\vec{e}_{ij}$ when $\vec{\tau}_i$ is aligned along  $\vec{e}_{ij}$,  where  $\chi_{ij}^u$ plays a role in the unit Finsler length, and hence, a large $\chi_{ij}^u$ means  a short Finsler length. As a consequence, this alignment of $\vec{\tau}_i$ along $\vec{e}_{ij}$ effectively makes diffusion coefficients $D_u$ direction dependent, which is numerically confirmed in the following section. This direction dependence of $D_u$ is observed in the model of Section \ref{Std-approach} when the parameter $a$ is increased in Eq. (\ref{direction-dep-diff-constants}). In contrast, the definition of $\chi_{ij}^v$ implies that an alignment of $\vec{\tau}_i$ along $\vec{e}_{ij}$ effectively makes the $D_v$ direction dependent, corresponding to the anisotropic TPs in Fig. \ref{fig-3}(f), (g) obtained when $b$ is decreased. Thus, we expect that these $\chi_{ij}^u$ and $\chi_{ij}^v$ correspond to the condition $a\!>\!b(=\!1)$ ($\Leftrightarrow c_v^u\!>\!1$) or $(1\!=)a\!<\!b$ ($\Leftrightarrow c_v^u\!<\!1$) shown in Fig. \ref{fig-4}{(a), (b) if $\vec{\tau}_i$ is aligned along $\vec{e}_{ij}$. Therefore, we expect that if the alignment of $\vec{\tau}_i$ is controlled to be aligned along the $x$ direction for almost all $i$ for example, then $\chi_{ij}^u$ and $\chi_{ij}^v$ in Eq. (\ref{Finsler-unit-lengths}) effectively plays a role in $(a,b)$ in Eq. (\ref{direction-dep-diff-constants}).

The discrete Laplace operators on triangulated lattices are obtained by the discrete $S_{u}$ and $S_{v}$ in Eq. (\ref{discrete-Hamiltonian}) such that
\begin{eqnarray}
\label{discrete-Laplace}
\begin{split}
&\Laplace u_i = 2\left(\sum_{j(i)} \gamma_{ij}^u u_j -u_i\sum_{j(i)} \gamma_{ij}^u\right), \\
&\Laplace v_i = 2\left(\sum_{j(i)} \gamma_{ij}^v v_j -u_i\sum_{j(i)} \gamma_{ij}^v\right)
\end{split}
\end{eqnarray}
for $1\!\leq\!i\!\leq\!N$, where $\sum_{j(i)}$ denotes the sum over vertices $j$ connected to $i$ (Appendix \ref{App-C}). Note that these Laplace operators are independent of the anisotropy parameters $a$ and $b$ in contrast to those in Eq. (\ref{discrete-Laplace-on-sq-latt}).

\subsection{Hybrid numerical technique \label{hybrid-tech}}
We introduce a numerical technique to find steady state configurations of $u$ and $v$ under the presence of IDOF $\tau$ on two types of triangulated lattices: fixed and fluid. The variables $u_i$ and $v_i$ are updated such that
\begin{eqnarray}
\label{discrete-t-iterations-tlattice}
\begin{split}
&u_{i}(t+{\Delta} t)\leftarrow u_{i}(t) +{\Delta} t \left(D_u{\Laplace}u_{i}(t) +f(u_{i}(t),v_{i}(t) \right),\\
&v_{i}(t+{\Delta} t)\leftarrow v_{ij}(t) +{\Delta} t \left(D_v{\Laplace}v_{i}(t) +g(u_{i}(t),v_{i}(t)\right), 
\end{split}
\end{eqnarray}
where the diffusion terms are given by Eq. (\ref{discrete-Laplace}). Direction dependence is not manually introduced in these diffusion terms, in sharp contrast to those of Eq. (\ref{discrete-Laplace-on-sq-latt}).
${\rm Max}\left\{|u_{i}(t\!+\!\Delta t)\!-\!u_{i}(t)|\right\} \!<\! 1\!\times\! 10^{-8}$ and
${\rm Max}\left\{|v_{i}(t\!+\!\Delta t)\!-\!v_{i}(t)|\right\} \!<\! 1\!\times\! 10^{-8}$, $(1\!\leq\!i\leq\!N)$,
which are the same relations as those in Eq. (\ref{convergence-std}) for the standard model on regular square lattices.

The hybrid numerical technique is summarized in the following four steps:
\begin{enumerate}
\item[(i)]  Discrete time evolution of Eq. (\ref{discrete-t-iterations-tlattice}) is iterated with the diffusion terms $\Laplace u_i$ and $\Laplace v_i$ in Eq. (\ref{discrete-Laplace}) with the coefficients $D_u$ and $D_v$.
\vspace{-2mm}
\item[(ii)] In each discrete time step $t \to t+\Delta t$, the variables $\{\vec{r}_i\}$ are updated once in MC simulations using $S$ in Eq. (\ref{discrete-total-Hamiltonian}), as shown in Fig. \ref{fig-4}(c), on a fixed connectivity lattice (fixed model) or on a dynamically triangulated fluid lattice (fluid model), which are defined by the partition functions in Eq. (\ref{part-funct}). At each update of $\vec{r}_i$, the variable $\vec{\tau}_i$ is also updated by using $\Delta \vec{r}_i(=\vec{r}_i({\rm new})\!-\!\vec{r}_i({\rm old})$ (Fig. \ref{fig-7}(c)).
\vspace{-2mm}
\item[(iii)] Steps (i) and (ii) are repeated $n_{\rm MC}$ times, where $n_{\rm MC}$ is suitably large.
\item[(iv)] Steps (i) are repeated under the final configurations of $\{\vec{\tau}_i\}$ in (iii) until the convergent criteria described above are satisfied.
\end{enumerate}

\section{Results \label{results}}
As described in the preceding section, the computational domain is considered to be spanned by a membrane with a surface tension $\sigma$, which depends on the lattice spacing $d$ (Fig. \ref{fig-6}(a)). For this reason, to check whether the results are influenced by $\sigma$ or not, we use two different values of $d$ such that
\begin{eqnarray}
\label{lattice-spacing}
\begin{split}
&d=0.525\quad(\Leftrightarrow \langle\ell^2\rangle\!\simeq\!\frac{1}{2}\Leftrightarrow \sigma>0), \\
&d=0.41\quad(\Leftrightarrow \langle\ell^2\rangle\!\simeq\!\frac{1}{3}\Leftrightarrow \sigma\simeq 0).
\end{split}
\end{eqnarray}
The mean squared bond lengths $\langle\ell^2\rangle$ inside the parenthesis are numerically obtained under the assumed $d$. Therefore, from the expression of surface tension
$\sigma\!=\!({3N}/{A})\left(\langle\ell^2\rangle \!-\!\frac{1}{3}\right)$ in Eq. (\ref{surface-tension}), we expect that $\sigma \! >\!0$ for $d\!=\!0.525$ and $\sigma \!\simeq\!0$ for $d\!=\!0.41$. We simply expect that the results are independent of $\sigma$ because $\sigma$ has no direction dependence.

\subsection{Snapshots \label{snapshot}}
The final configurations of $u$ and $v$ depend on their initial configurations and on $\tau$, and the obtained snapshots are almost identical but not always exactly the same if the initial values of these variables are different from each other. For this reason, steps (i) to (iv) described in Section \ref {hybrid-tech} are iterated once in the simulations to obtain snapshots in this subsection.

Figure \ref{fig-9}(a)--(d) show snapshots of the variable $u$ without the IDOF $\vec{\tau}$ (upper row) and with $\vec{\tau}$ (lower row) of the fixed model on the $N\!=\!10000$ lattice and the lattice spacing $d\!=\!0.525$. The values of $u$ are normalized in the range $[0,1]$ in the snapshots, as shown in the color code. These snapshots are obtained by varying $\lambda$ and $\vec{F}$, both of which influence the alignment of $\vec{\tau}$; $\lambda$ is the strength of the nearest neighbor correlation of the IDOF $\vec{\tau}$, and $\vec{F}$ aligns $\vec{\tau}$ along $\vec{F}$. The other parameters are fixed to $(\alpha,\gamma)\!=\!(1,8)$, and $(D_u,D_v)\!=\!(0.4,5)$. The $\chi_0$ in Eq. (\ref{Finsler-unit-lengths}) is assumed to be $\chi_0\!=\!0.5$.
When $\lambda\!=\!0$ and $\vec{F}\!=\!(0,0)$, the vertex positions $\vec{r}$ are expected to be random, causing a random fluctuation of IDOF $\vec{\tau}$. In this case, an isotropic pattern is expected. These expectations are confirmed in Fig. \ref{fig-9}(a), where the variables $\vec{\tau}$ in the snapshots of the central region are plotted with small cones to clarify their directions, although $\vec{\tau}$ and $-\vec{\tau}$ are identified in $S_\tau$ and $S_\tau^F$ of Eq. (\ref{discrete-Hamiltonian}) and in $\nu_{ij}^u$ and $\nu_{ij}^v$ of Eq. (\ref{Finsler-unit-lengths}). When $\vec{F}$ is increased to $\vec{F}\!=\!(1,0)$ (Fig. \ref{fig-9}(b)), the pattern becomes aisotropic, and $\vec{\tau}$ is also slightly aligned along the $x$ direction. The alignment of $\vec{\tau}$ becomes clear when $\lambda$ is increased to $\lambda\!=\!0.5$ (Fig. \ref{fig-9}(c)), and the pattern is also forced to be more anisotropic by enlarging $\vec{F}\!=\!(0,0)$ to $\vec{F}\!=\!(0,2)$, where the direction is changed to the $y$-axis. From these snapshots, we find that the isotropy/anisotropy in the patterns is determined by the fluctuation direction $\tau$ of the vertices, which is controlled by the external force $\vec{F}$. This finding implies that the FG modeling prescription effectively modifies the diffusion constants $D_u$ and $D_v$ in an anisotropic or direction-dependent manner. These effective diffusion constants are numerically extracted from the simulation data in the following subsection.
Notably, the patterns of $v$ are almost the same as those of $u$ plotted in Fig. \ref{fig-9}.

\begin{figure}[h!]
\centering{}\includegraphics[width=14.5cm]{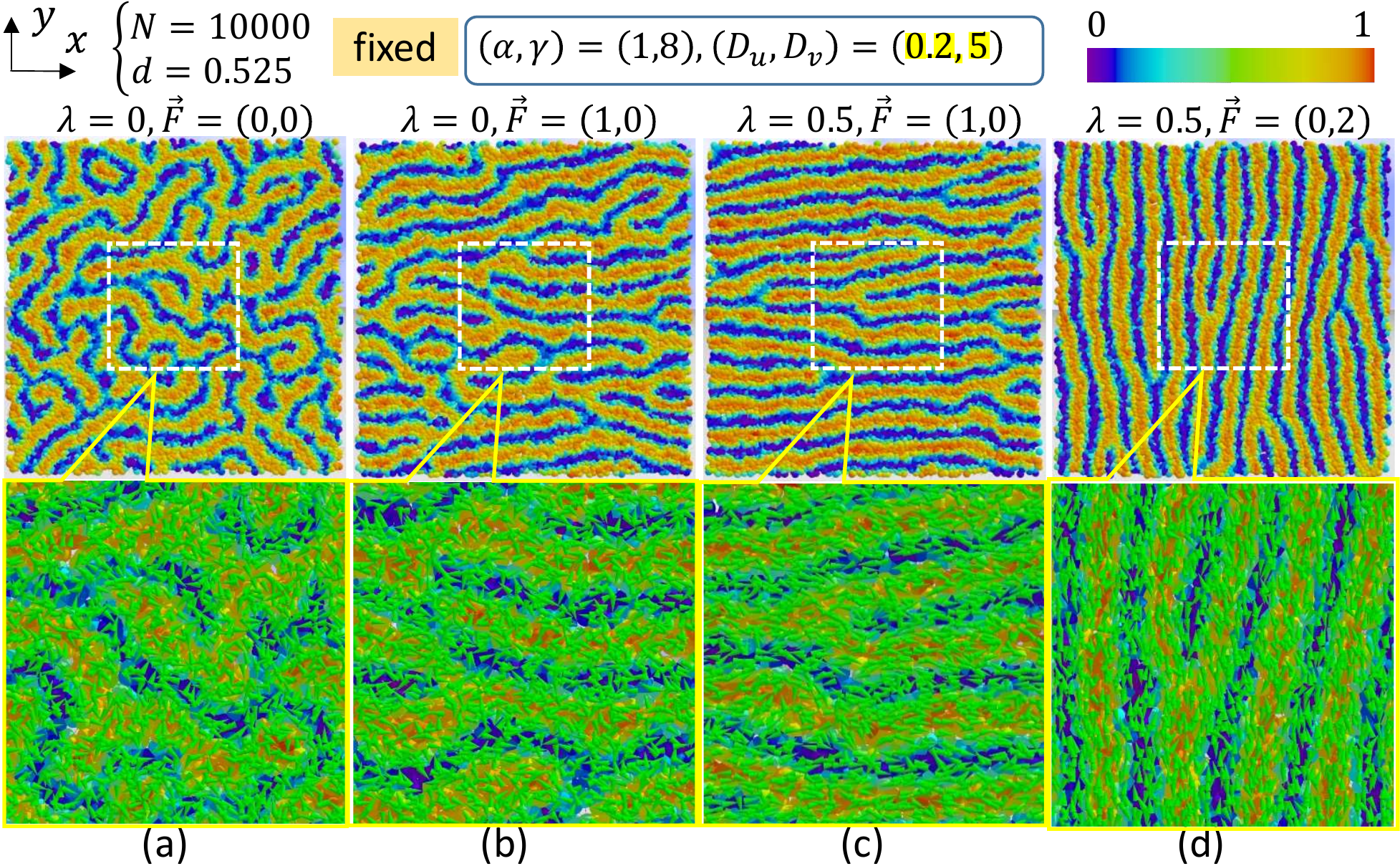}
\caption {Snapshots of TPs obtained on the fixed lattice of size $N\!=\!10000$ with lattice spacing $d\!=\!0.525$. The assumed parameters are $(\alpha,\gamma)\!=\!(1,8)$ and $(D_u,D_v)\!=\!(0.2,5)$, with (a) $\lambda\!=\!0, \vec{F}\!=\!(0,0)$, (b) $\lambda\!=\!0, \vec{F}\!=\!(1,0)$, (c) $\lambda\!=\!0.5, \vec{F}\!=\!(1,0)$, and (d) $\lambda\!=\!0.5, \vec{F}\!=\!(0,2)$. The upper rows show the normalized $u (\in [0,1])$ indicated by the color code without the IDOF $\vec{\tau}$, and the lower row shows enlarged views with $\vec{\tau}$ of the central region enclosed by the dashed rectangles, in which $\vec{\tau}$ are represented by small cones.
\label{fig-9}}
\end{figure}
The patterns obtained with $d\!=\!0.41$ corresponding to $\sigma\!\simeq\!0$ in Eq. (\ref{lattice-spacing}) are almost the same as those in Fig. \ref{fig-9} obtained with $d\!=\!0.525$, and therefore, we find that the patterns are not influenced by $\sigma$, as expected. The fact that the results are independent of $\sigma$ is natural because $\sigma$ is isotropic (Fig. \ref{fig-A-1}) and does not change the lattice structure anisotropy. The pattern anisotropy is caused by direction-dependent alignment of $\vec{\tau}$ reflecting anisotropy in the vertex fluctuations. Influences of $\sigma$ on TPs are expected in the case of anisotropic $\sigma$. This expectation is discussed in the following subsection.

\begin{figure}[h!]
\centering{}\includegraphics[width=14.5cm]{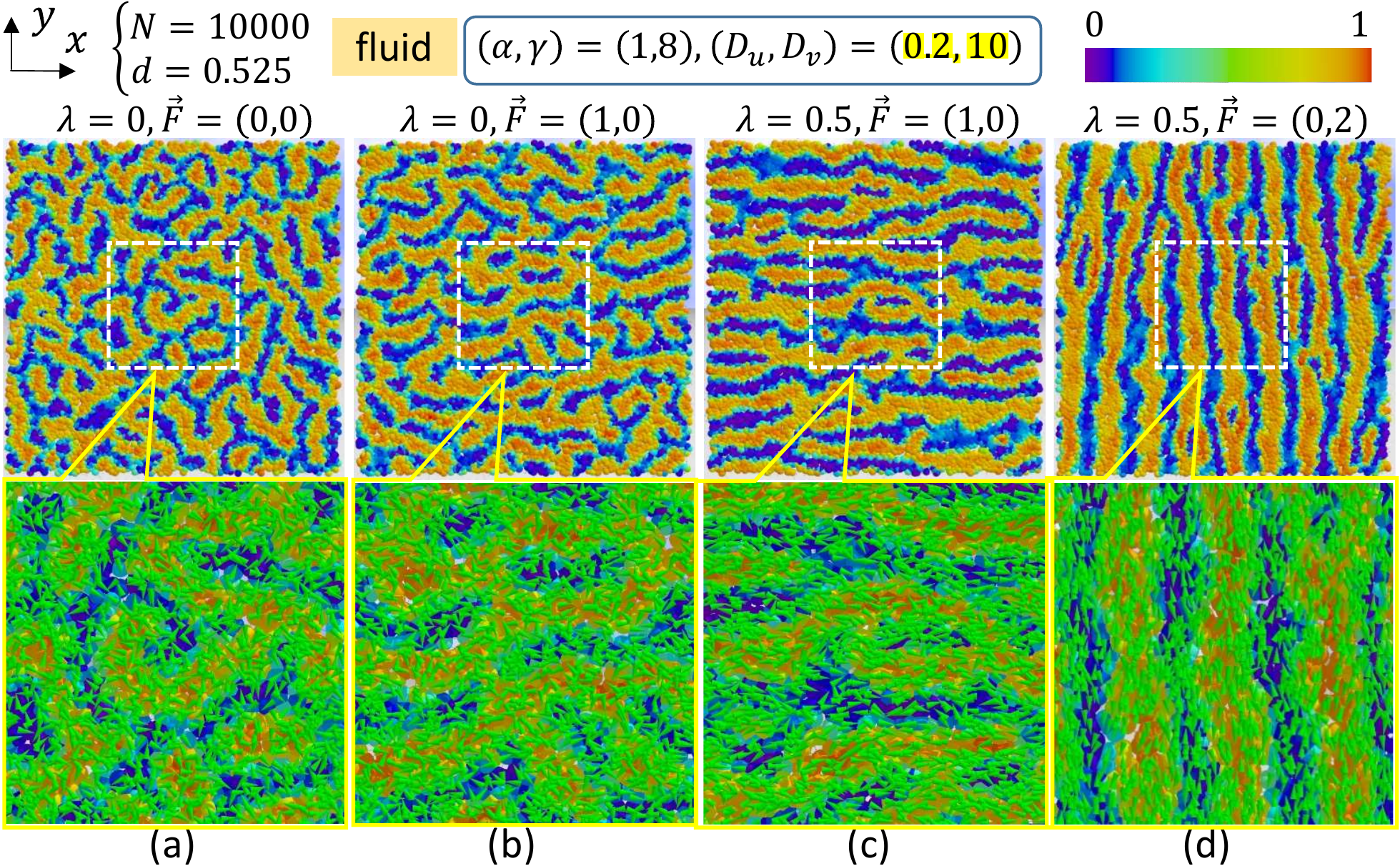}
\caption {Snapshots of TPs obtained on the fluid lattice of size $N\!=\!10000$ and lattice spacing $d\!=\!0.525$. The assumed parameters are $(\alpha,\gamma)\!=\!(1,8)$ and $(D_u,D_v)\!=\!(0.2,10)$, with (a) $\lambda\!=\!0, \vec{F}\!=\!(0,0)$, (b) $\lambda\!=\!0, \vec{F}\!=\!(1,0)$, (c) $\lambda\!=\!0.5, \vec{F}\!=\!(1,0)$, and (d) $\lambda\!=\!0.5, \vec{F}\!=\!(0,2)$. These are the same as those assumed in Fig. \ref{fig-9} except for $D_v\!=\!10$, which is twice as large as that of Fig. \ref{fig-9}.
\label{fig-10}}
\end{figure}
Snapshots obtained on the fluid lattice are shown in Fig. \ref{fig-10}(a)--(d), where the parameters are the same as those for the fixed model presented in Fig. \ref{fig-9} except $D_v\!=\!10$, which is $D_v\!=\!5$ in Fig. \ref{fig-9}. If $D_v\!=\!5$ is used for the fluid model, some of the patterns in Fig. \ref{fig-10} do not appear. This difference is considered to come from a difference in lattice structure between fixed and fluid lattices. The vertex positions $\vec{r}$ on fluid lattices are expected to be more influenced by the external force $\vec{F}$ and their nearest neighbors via the correlation energy $S_\tau$ due to the free diffusion of vertices shown in Fig. \ref{fig-8}(c). To see such a difference in the lattice structure, we calculate the mean value distribution of the bond length $h(\ell)$ as well as the direction-dependent diffusion constants in the following subsection.

\subsection{Direction-dependent diffusion constants and surface tension \label{surface-tension}}
In this subsection, we plot direction-dependent diffusion constants (Appendix \ref{App-D})
\begin{eqnarray}
\label{effective-diffusion-constants-phsi}
\begin{split}
&D_{u}^x=\frac{1}{N_B}\sum_{ij}\gamma_{ij}^{u}\cos^2\theta_{ij}, \quad
D_{u}^y=\frac{1}{N_B}\sum_{ij}\gamma_{ij}^{u}\sin^2\theta_{ij}, \\
&D_{v}^x=\frac{1}{N_B}\sum_{ij}\gamma_{ij}^{v}\cos^2\theta_{ij}, \quad
D_{v}^y=\frac{1}{N_B}\sum_{ij}\gamma_{ij}^{v}\sin^2\theta_{ij},
\end{split}
\end{eqnarray}
and the corresponding direction-dependent energies, where $N_B(=\!3N)$ is the total number of bonds and $\theta_{ij}$ is the angle between $\vec{e}_{ij}$ and the $x$-axis (Appendix \ref{App-D}),
\begin{eqnarray}
\label{effective-energy-phsi}
\begin{split}
&S_u^x=\frac{1}{\langle\gamma_{ij}^u\cos^2\theta_{ij}\rangle}\sum_{ij} \gamma_{ij}^u\cos^2\theta_{ij}(u_i-u_j)^2, \;
S_u^y=\frac{1}{\langle\gamma_{ij}^u\sin^2\theta_{ij}\rangle}\sum_{ij} \gamma_{ij}^u\sin^2\theta_{ij}(u_i-u_j)^2,\\
&S_v^x=\frac{1}{\langle\gamma_{ij}^v\cos^2\theta_{ij}\rangle}\sum_{ij} \gamma_{ij}^v\cos^2\theta_{ij}(v_i-v_j)^2, \;
S_u^y=\frac{1}{\langle\gamma_{ij}^v\sin^2\theta_{ij}\rangle}\sum_{ij} \gamma_{ij}^v\sin^2\theta_{ij}(v_i-v_j)^2.
\end{split}
\end{eqnarray}
We calculate these quantities using the final configurations obtained at step (iv) described in Section \ref{hybrid-tech}. The final configurations of $u$, $v$ and $\tau$ depend on their initial configurations, and for this reason, we obtain their mean final values by repeating steps (i) to (iv) $n_{\rm itr}$ with random initial configurations. The $n_{\rm itr}$, the total number of iterations $n_{\rm MC}$ of steps (i) and (ii), and the lattice size $N$ for the simulations in this subsection are assumed as follows:
\begin{eqnarray}
\label{nitr-and-N}
n_{\rm itr}=50,\quad n_{\rm MC}=5\times 10^5,\quad N=6400.
\end{eqnarray}

Figure \ref{fig-11}(a) shows $D_u^\mu$ vs. $F$ and $S_u^\mu/N_B$ vs. $F$, where $\vec{F}\!=\!(F,0)$ and $N_B$ is the total number of bonds. Open and solid symbols correspond to the data obtained under $d\!=\!0.525$ and $d\!=\!0.41$, respectively, and the data are almost independent of the difference in the lattice spacing $d$.
We also find that $D_{u}^x$ ($D_{u}^y$) increases (decreases) with increasing $F$. Note that the IDOF $\vec{\tau}$ is almost random at $F\to 0$ and aligns along the $x$-axis when $F$ increases. This behavior of $D_{u}^x$ and $D_{u}^y$ in Fig. \ref{fig-11}(a) is consistent with those $D_{u}^x\!=\!aD_u$ and $D_{u}^y\!=\!(2-a)D_u$ expected when $a$ increases from $a\!=\!1$ in Eq. (\ref{direction-dep-diff-constants}) in the preceding section. Moreover, we also find that $S_u^x/N_B$ ($S_u^y/N_B$) decreases (increases) with increasing $F$. These results indicate that $S_u^x/N_B$ ($S_u^y/N_B$) decreases (increases) when $D_{u}^x$ increases ($D_{u}^y$ decreases). Here, we note that $S_u^\mu/N_B$ and $D_{u}^\mu$ have no relationship between input and output, but both $S_u^\mu/N_B$ and $D_{u}^\mu$ are outputs to the input $F$.
$D_v^\mu$ vs. $F$ and  $S_v^\mu/N_B$ vs. $F$ are  plotted in Fig.  \ref{fig-11}(b), in which we also find that the variation in $S_v^\mu/N_B$ is closely related with that of $D_{v}^\mu$.

\begin{figure}[h!]
\centering{}\includegraphics[width=11.5cm]{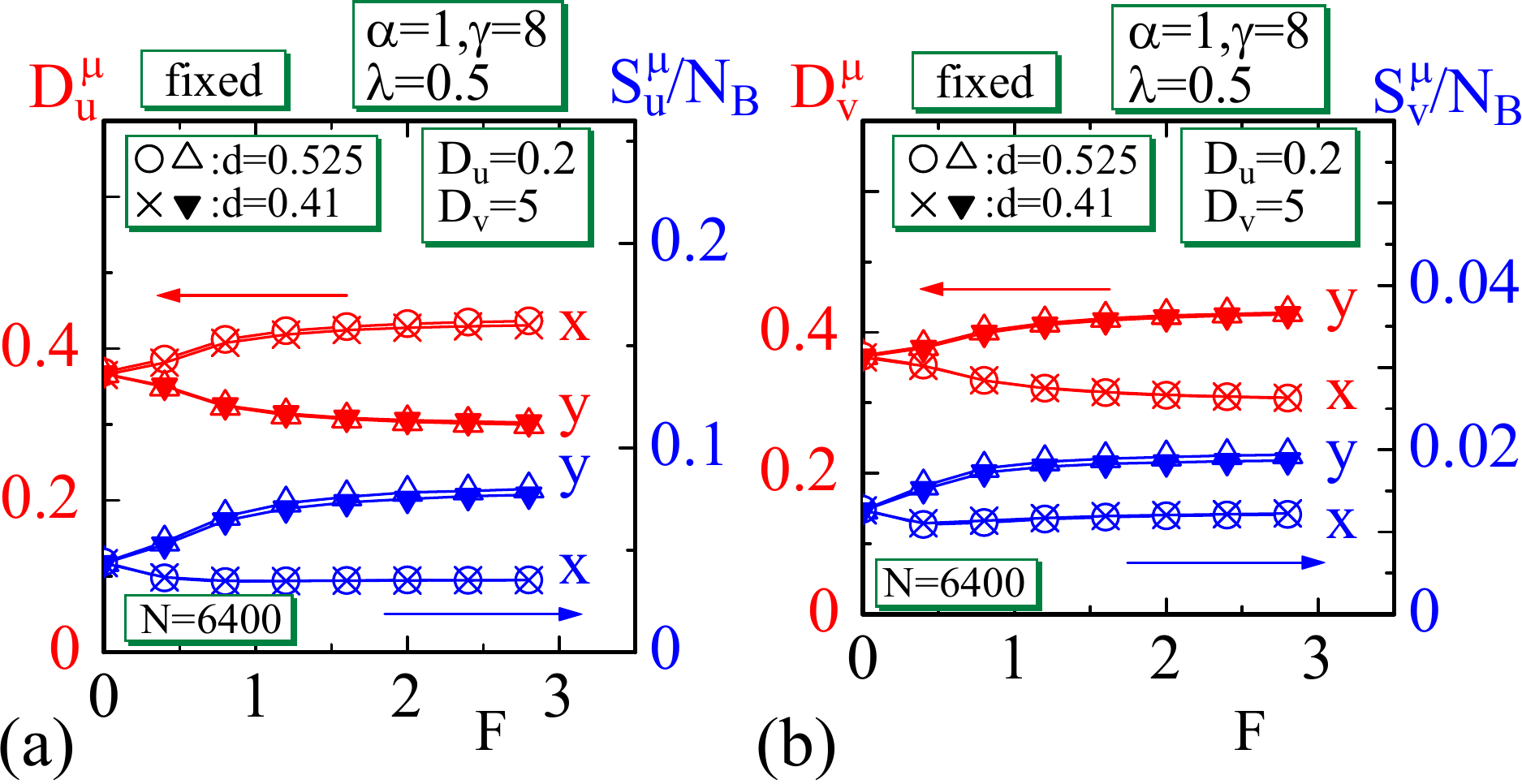}
\caption {(a) Diffusion constants $D_u^\mu$ in Eq. (\ref{effective-diffusion-constants-phsi}) and the direction-dependent energies $S_u^\mu$ vs. $F$, and (b) $D_v^\mu$ and $S_v^\mu$ vs. $F$ for the fixed model on a lattice of size $N\!=\!6400$. The data obtained under the lattice spacing $d\!=\!0.525$ ($d\!=\!0.41$) are denoted by the open (solid) symbols. The results are almost independent of $d$.
The total number of iterations is set to $n_{\rm itr}\!=\!50$ at each $F$ to calculate the mean values of these physical quantities.
\label{fig-11}}
\end{figure}
To compare the results plotted in Fig. \ref{fig-11}(a), (b) with those in Fig. \ref{fig-4}(a), (b), we consider the results in Fig. \ref{fig-11}(a) and (b) as the responses $S_u^\mu$ and $S_v^\mu$ to the inputs $D_{u}^\mu$ and $D_{v}^\mu$ as follows:
\begin{eqnarray}
&&{\rm Fig. \ref{fig-11}(a):}\; S_u^x \searrow \;({\rm if}\; D_u^x\nearrow), \quad S_u^y  \nearrow \;({\rm if}\; D_u^y\searrow),  \label{results-FG-1} \\
&&{\rm Fig. \ref{fig-11}(b):}\; S_v^x \searrow \;({\rm if}\; D_v^x\searrow), \quad S_v^y  \nearrow \;({\rm if}\; D_v^y\nearrow). \label{results-FG-2}
\end{eqnarray}
Thus, we find that the results in Eqs. (\ref{results-FG-1}) and (\ref{results-FG-2}) are consistent with the behaviors of $d_\mu^2u$ and $d_\mu^2v$ vs. $a$ and $b$ in Fig. \ref{fig-4}(a) and (b). Indeed, the descriptions in Eq. (\ref{results-FG-1}) are consistent with the first line in Eq. (\ref{results-std-1}), and those in Eq. (\ref{results-FG-2}) are consistent with the third line in Eq. (\ref{results-std-2}), although $S_u^\mu/N_B$ ($S_v^\mu/N_B$) does not directly correspond to $d_{\mu}^2u$ ($d_{\mu}^2v$). However, it is natural to expect that variation of $D_u^\mu$ ($D_v^\mu$) included in $S_u\!=\!D_u^x S_u^x\!+\!D_u^x S_u^y$ ($S_v\!=\!D_v^x S_v^x\!+\!D_v^x S_v^y$) makes the remaining part $S_u^\mu$ ($S_v^\mu$) vary oppositely in MC updates owing to the expectation that $S_u$ ($S_v$) is almost at equilibrium and hence almost stable.
For the same reason, the coefficients $a$, $2\!-\!a$ and  $b$, $2\!-\!b$ in the diffusion terms  ${\Laplace}_au_{ij}$ and  ${\Laplace}_bv_{ij}$ in Eq. (\ref{discrete-Laplace-on-sq-latt}) influence the corresponding second-order partial differentials $\partial^2_{x,y} u$ and $\partial^2_{x,y}v$, variations of which are reflected in $d_\mu^2u$ and $d_\mu^2v$. This consistency between Eqs. (\ref{results-FG-1}), (\ref{results-FG-2}) and Eqs. (\ref{results-std-1}), (\ref{results-std-2}) implies that the FG modeling with the IDOF $\tau$ suitably implements the diffusion anisotropy in TPs.
The same consistency is obtained with the results on the fluid lattice plotted in Fig. \ref{fig-12}(a) and (b).
\begin{figure}[h!]
\centering{}\includegraphics[width=11.5cm]{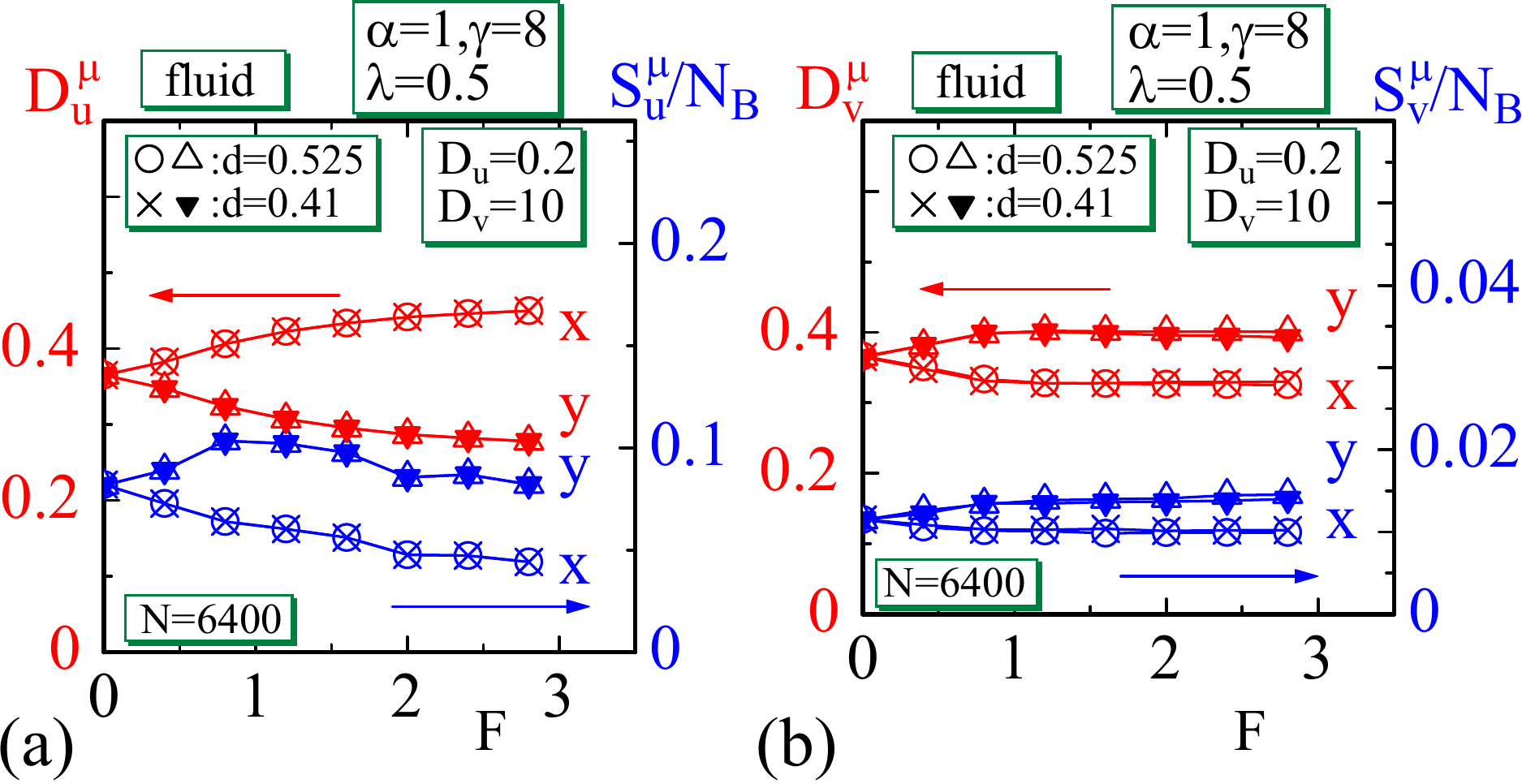}
\caption {(a) Diffusion constants $D_u^\mu$ in Eq. (\ref{effective-diffusion-constants-phsi}) and direction-dependent energies $S_u^\mu$ vs. $F$ and (b) $D_v^\mu$ and $S_v^\mu$ vs. $F$ for the fluid model on a lattice of size $N\!=\!6400$. The data obtained under the lattice spacing $d\!=\!0.525$ ($d\!=\!0.41$) are denoted by the open (solid) symbols, indicating no dependence of the results on $d$.
\label{fig-12}}
\end{figure}

\begin{figure}[h!]
\centering{}\includegraphics[width=11.5cm]{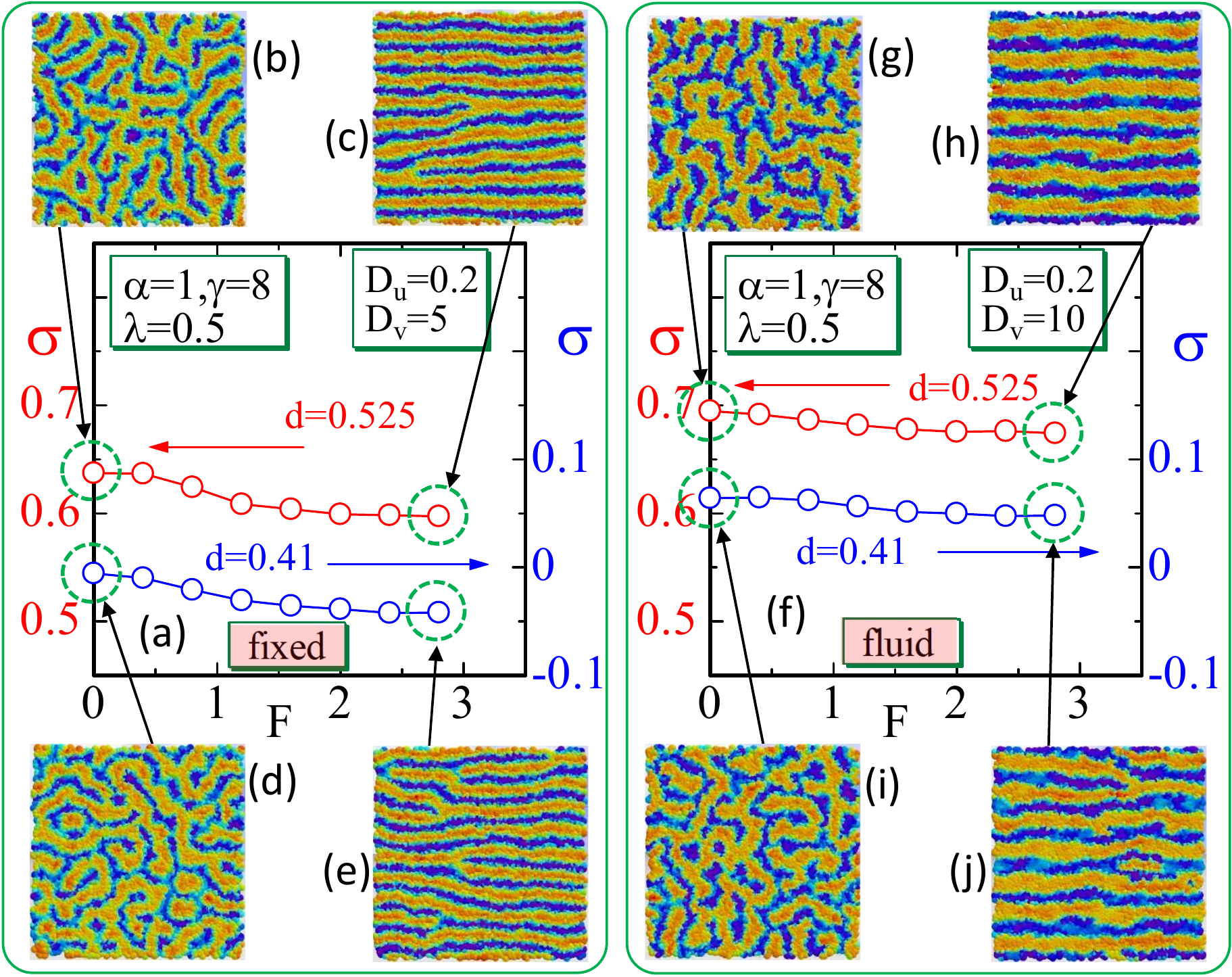}
\caption {Surface tension $\sigma$ vs. $F$ ($\vec{F}\!=\!(F,0)$) on the (a) fixed lattice with (b)--(e) snapshots of $u$ at $F\!=\!0$ and $F\!=\!2.8$ and on the (f) fluid lattice with (g)--(j) snapshots of $u$ at $F\!=\!0$ and $F\!=\!2.8$, where the lattice spacing is assumed to $d\!=\!0.525$ and $d\!=\!0.41$. The lattice size ($N\!=\!6400$) and the parameters are the same as those assumed in Figs. \ref{fig-11} and \ref{fig-12}. We find that $\sigma(d\!=\!0.525)>0$ and $\sigma(d\!=\!0.41)\simeq 0$, and both of them decrease with increasing $F$.
\label{fig-13}}
\end{figure}
Next, we show the surface tension $\sigma$ in Eq. (\ref{surface-tension}) for the lattice spacing $d\!=\!0.525$ and $d\!=\!0.41$ in Fig. \ref{fig-13}(a) and (b). We have checked in the preceding subsections that the patterns are only minimally influenced by $\sigma$; however, it is still interesting to see interaction between $\sigma$ and IDOF $\tau$, which is controlled by the external force $\vec{F}$.
We find from Fig. \ref{fig-13}(a),(b) that $\sigma(d\!=\!0.525)>0$ and $\sigma(d\!=\!0.41)\simeq 0$ as expected, and moreover, $\sigma$ slightly decreases as $F$ increases in both fixed and fluid lattices. A decrease in $\sigma$ corresponds to a decrease in the mean bond length-squares $\langle \ell^2\rangle$ for $\sigma(d\!=\!0.525)$ and $\sigma(d\!=\!0.41)$ because of the expression of $\sigma\propto d^{-2}(\langle\ell^2\rangle \!-\!{1}/{3})$ in Eq. (\ref{surface-tension}) (see Appendix \ref{App-E} for the distribution of $\ell^2$ on both fixed and fluid lattices.)
This decrease in $\langle \ell^2\rangle$ indicates that the vertex distribution becomes uniform in the large-$F$ region compared with a nonuniform distribution, where a larger $\langle \ell^2\rangle$ is expected because the vertices are confined in the fixed area $A\!=\!(n_x\!-\!1)(n_y\!-\!1)d^2$, where $n_x\!=\!n_y\!=\!80, d\!=\!0.525$. This dependence of $\sigma$ on $F$ implies that $\sigma$ depends on IDOF $\tau$ because $F$ directly influences $\tau$, and therefore, we consider that $\tau$ interacts with $\sigma$. This interaction between $\tau$ and $\sigma$, which is controlled by a frame or boundary conditions, allows us to control $\tau$ with $\sigma$. As mentioned in the first part of this section, the surface tension $\sigma$ imposed by PBCs in both $x$ and $y$ directions are isotropic and does not deform the lattice except for the uniform or isotropic expansion/compression, and hence, a nontrivial effect is not expected in the TPs. However, if $\sigma$ is anisotropic, $\sigma$ influences the IDOF $\tau$. Such an anisotropic $\sigma$ is expected to be caused by direction-dependent boundary conditions. Therefore, $\vec{\tau}$ can be controlled by mechanical boundary conditions as well as the external force $\vec{F}$. This prediction is confirmed in the following subsection.

Notably, $\sigma$ decreases as the TPs change from isotropic to anisotropic. A small $\sigma$ comes from the small tensile energy $S_1$ in Eq. (\ref{discrete-Hamiltonian}), implying that anisotropic TPs are mechanically stable. Anisotropic TPs are caused by an alignment of IDOF $\vec{\tau}$ corresponding to lower energy configurations with respect to $S_{\tau}$ in Eq. (\ref{discrete-Hamiltonian}). Thus, we find that the alignment of $\vec{\tau}$ lowers the mechanical energy $S_1$ as well as $S_{\tau}$. This interaction between $\tau$ and $\sigma$ implies that anisotropy in the fluctuation directions of $\vec{r}$ influences the position of $\vec{r}$ because $\tau$ represents the fluctuation directions of $\vec{r}$. Therefore, FG modeling based on IDOF $\vec{\tau}$ is considered reasonable.

\subsection{Control of the pattern direction \label{direction-control}}
\begin{figure}[t!]
\centering{}\includegraphics[width=11.5cm]{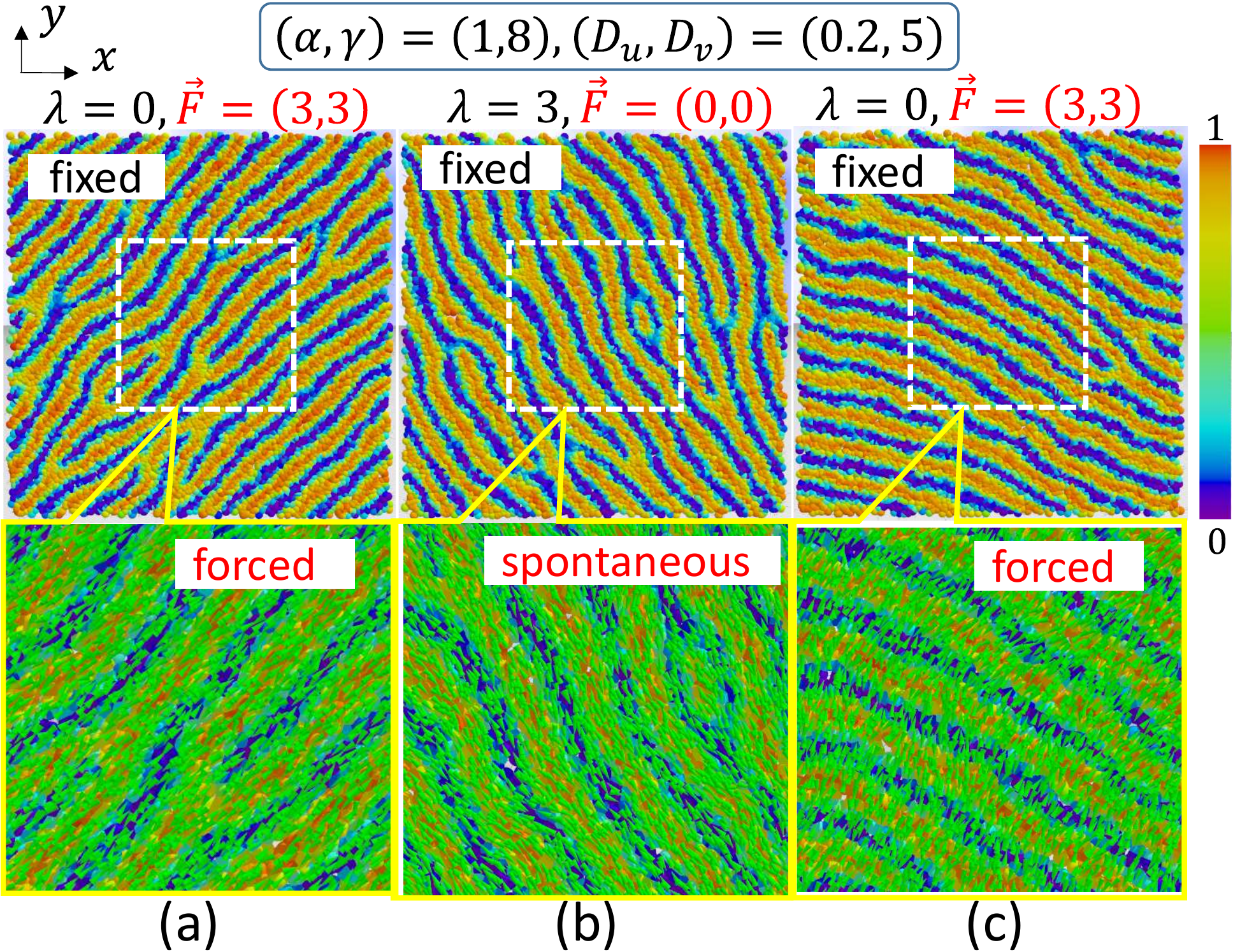}
\caption {Snapshots of TPs, of which the direction is (a) ``forced'' by external $\vec{F}$ and (b) spontaneously determined without $\vec{F}$. The central domains are plotted with small cones representing IDOF $\vec{\tau}$ in the lower part; (c) ``forcing'' with the replacement of $\chi_{ij}^u \leftrightarrow \chi_{ij}^v$ in Eq. (\ref{Finsler-unit-lengths-App}), where the pattern direction is vertical to IDOF $\vec{\tau}$.
\label{fig-14}}
\end{figure}
First, we show that the pattern direction can be arbitrarily and spontaneously determined. Snapshots in Fig. \ref{fig-14}(a) denoted by ``forced'' are obtained with $\vec{F}\!=\!(3,3)$ and $ \lambda\!=\!0$. In this case, the direction of IDOF $\tau$ is controlled by $\vec{F}$, and consequently, patterns align along $\tau$. On the other hand, snapshots in Fig. \ref{fig-14}(b) denoted by ``spontaneous'' are obtained under $\vec{F}\!=\!(0,0)$ and $\lambda\!=\!3$. In this case, IDOF $\vec{\tau}$ aligns to a spontaneously determined direction due to a relatively large $\lambda(=\!3)$; pattern directions depend on random numbers for initial random configurations of $\vec{\tau}$, for instance. These properties on pattern directions, forced and spontaneous, are specific to FG models. Figure \ref{fig-14}(c) is obtained by ``forcing'' with $\vec{F}\!=\!(3,3)$, the same as in Fig. \ref{fig-14}(a); however, the pattern direction is almost vertical to that in Fig. \ref{fig-14}(a). The reason is that $\chi_{ij}^u$ in Eq. (\ref{Finsler-unit-lengths-App}) is replaced by $\chi_{ij}^v$ such that $\chi_{ij}^v\!=\!|\vec{\tau}_i\cdot\vec{e}_{ij}| \!+\! \chi_0$ and $\chi_{ij}^u\!=\!\sqrt{1\!-\!|\vec{\tau}_i\cdot\vec{e}_{ij}|^2} \!+\! \chi_0$. In this case, $\vec{\tau}$ direction is perpendicular to the pattern direction as we confirm from the snapshot in the lower part of Fig. \ref{fig-14}(c).

\begin{figure}[t!]
\centering{}\includegraphics[width=11.5cm]{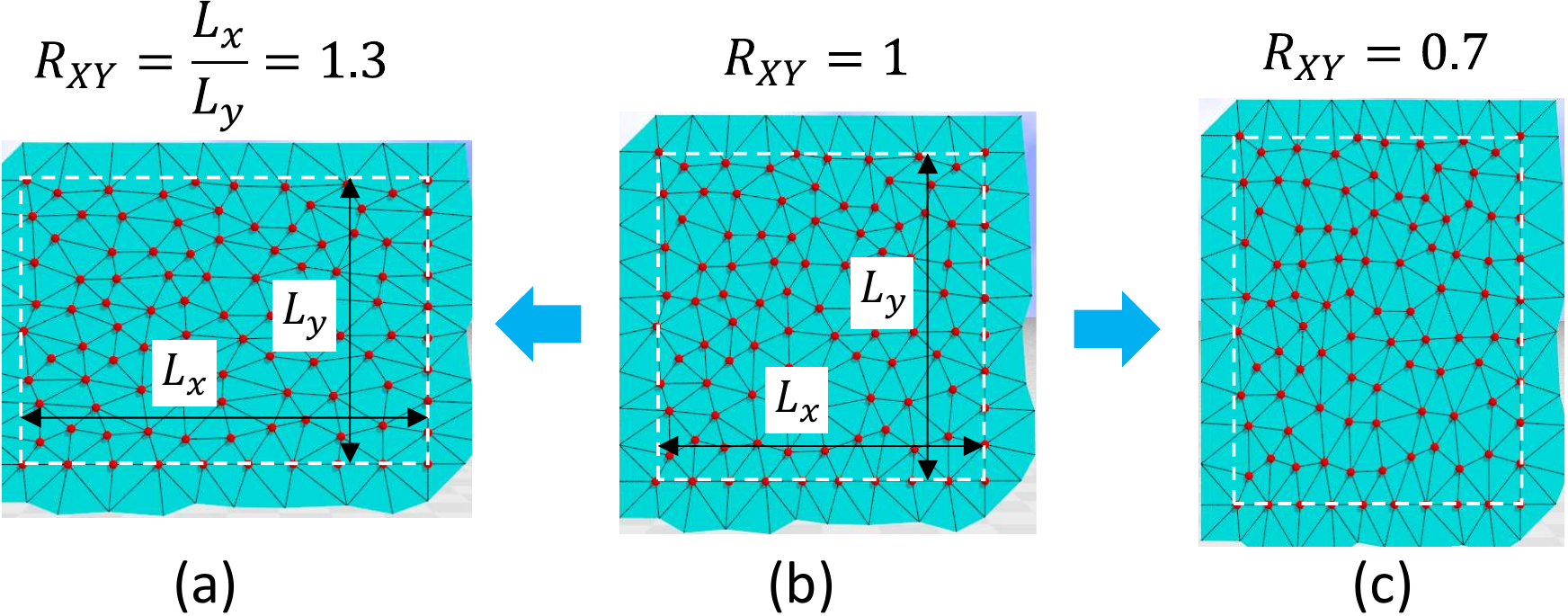}
\caption {Initial configurations of deformed and original lattices. Deformation does not change the lattice size $N\!=\!100$, which is very small for visualization of the lattice structure. The ratio of side lengths is (a) $R_{XY}\!=\!1.3$, (b) $R_{XY}\!=\!1$ and (ca) $R_{XY}\!=\!0.7$. The area $L_x L_Y$ also remains unchanged under deformation.
\label{fig-15}}
\end{figure}
Now, we deform the side lengths $L_x$ and $L_y(=\!L_x)$ of the lattice using the parameter $R_{XY}$ in $L_x^\prime\!=\!\sqrt{R_{XY}}L_x$, $L_y^\prime\!=\!L_y/\sqrt{R_{XY}}$ so that $R_{XY}\!=\!L_x^\prime/L_y^\prime$ (Fig. \ref{fig-15}(a)-(c)). Both the area $L_xL_y$ and the size $N$ remain unchanged under deformation.

\begin{figure}[t!]
\centering{}\includegraphics[width=13.5cm]{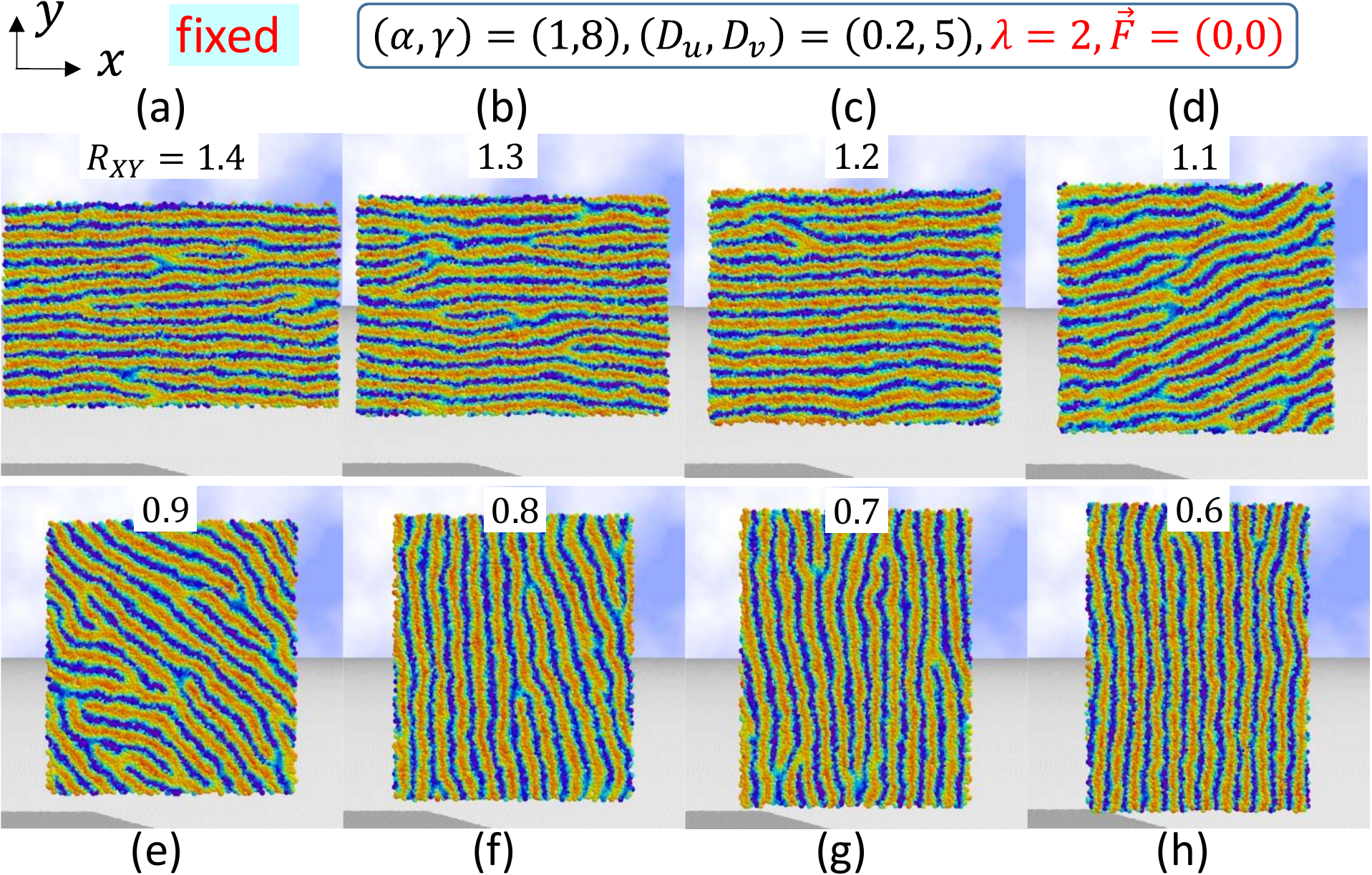}
\caption {Alignment of anisotropic TPs along the longer direction of lattices deformed with ratio $R_{XY}\!=\!L_x/L_y$ ranging from (a) $R_{XY}\!=\!1.4$ to (h) $R_{XY}\!=\!0.6$. These alignments are caused by strains applied on the boundary frame of lattices.
\label{fig-16}}
\end{figure}
Snapshots on deformed lattices with the range $1.4\!\geq\!R_{XY}\!\geq\!0.6$ of the fixed model are shown in Fig. \ref{fig-16}(a)-(h). The TP direction aligns along the longer direction if the ratio $R_{XY}$ deviates from $R_{XY}\!=\!1$ to a certain extent. The parameters are shown in the figure. The coefficients $\lambda$ and $\vec{F}$ are fixed to $\lambda\!=\!2$ and $\vec{F}\!=\!(0,0)$, and therefore, these alignments emerge because the spontaneous direction is determined uniquely by the lattice deformation. As shown in Fig. \ref{fig-15}(a), (c), the bond lengths along the longer direction are longer than those along the shorter direction. For this reason, vertices move along the longer direction relatively easily compared to the vertical direction, and therefore, IDOF $\vec{\tau}$ aligns along the longer direction. This is the alignment mechanism of anisotropic TPs due to the boundary condition. We note that the lattice deformation is caused by uniaxial tensile strains or compressions, and therefore, we consider that mechanical strains applied on the boundary impart TP anisotropy.

\begin{figure}[t!]
\centering{}\includegraphics[width=10.5cm]{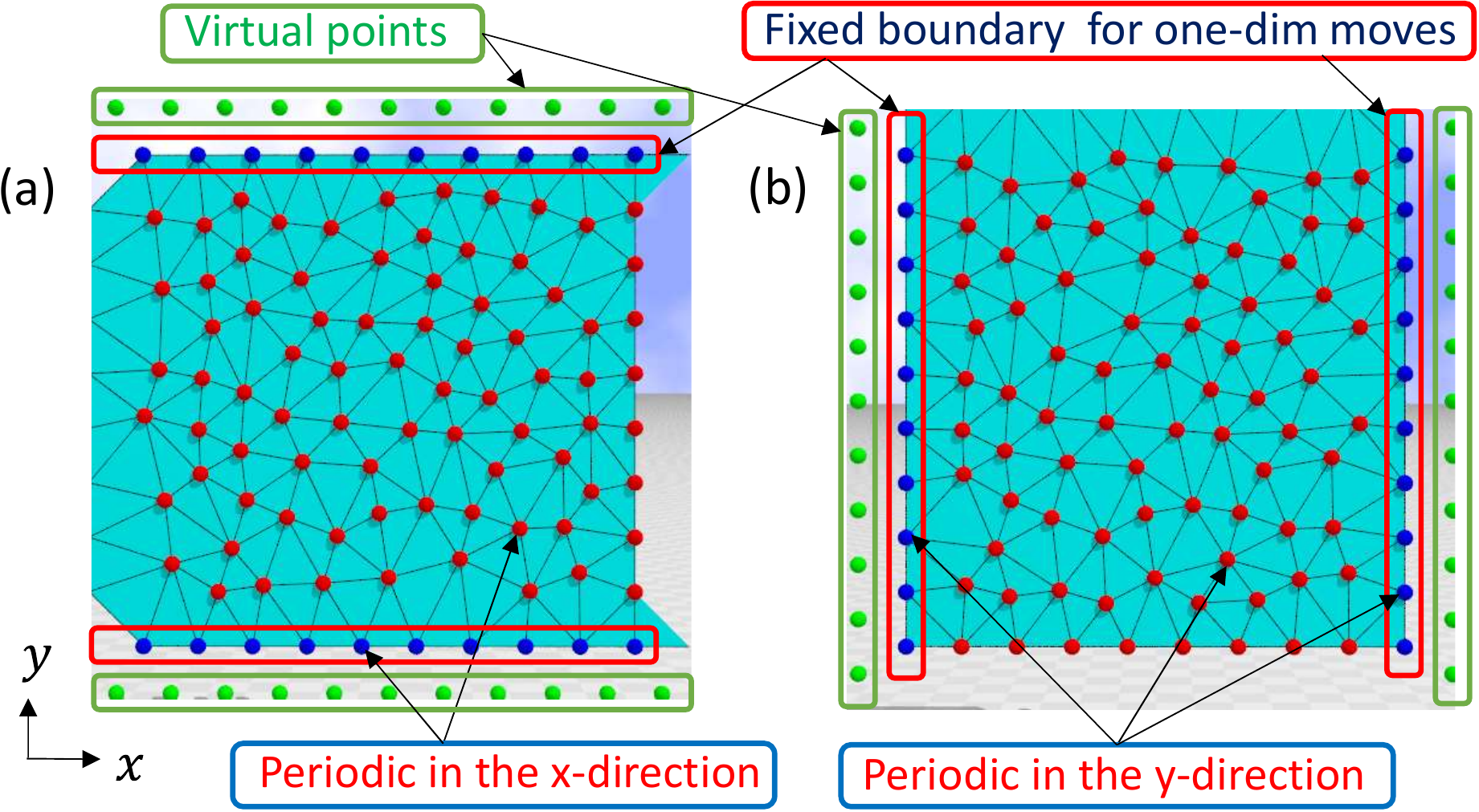}
\caption { (a) PBC is assumed in (a) the $x$ direction for vertices in red color (\textcolor{red}{$\bullet$}) and blue color (\textcolor{blue}{$\bullet$}) denoted by ``Fixed boundary for one-dim moves'', where the vertices can move along the boundary and the boundary bonds are not flipped (Fig. \ref{fig-7}(b)). The vertices  (\textcolor{green}{$\bullet$}) outside the fixed boundaries are ``virtual points'', which are necessary to perform Voronoi tessellation for triangulation and to define $\gamma_{ij}^{u,v}$ in Eq. (\ref{discrete-Hamiltonian}) on the boundaries. (b)  PBC is assumed in the $y$ direction, where the directions of the fixed boundary and virtual points are opposite to those in (a).  The lattice size is $N\!=\!100$, which is the sum of vertices (\textcolor{red}{$\bullet$}) and 
 (\textcolor{blue}{$\bullet$}). On the virtual vertices, the dynamical variables are not defined.
\label{fig-17}}
\end{figure}
On fluid lattices, this mechanism does not work because vertices freely move due to their fluid nature. Nevertheless, even on fluid lattices, vertices fluctuate along a direction in which PBCs are assumed, while the fixed boundary condition is assumed to prohibit vertices from undergoing free diffusion in the perpendicular direction. In this case, the spontaneous alignment direction of TPs is also expected to be determined by the boundary condition (Figs. \ref{fig-17}(a),(b)).
In this case, the spontaneous alignment direction of TPs is expected to be determined also by the boundary condition. 
\begin{figure}[t!]
\centering{}\includegraphics[width=10.5cm]{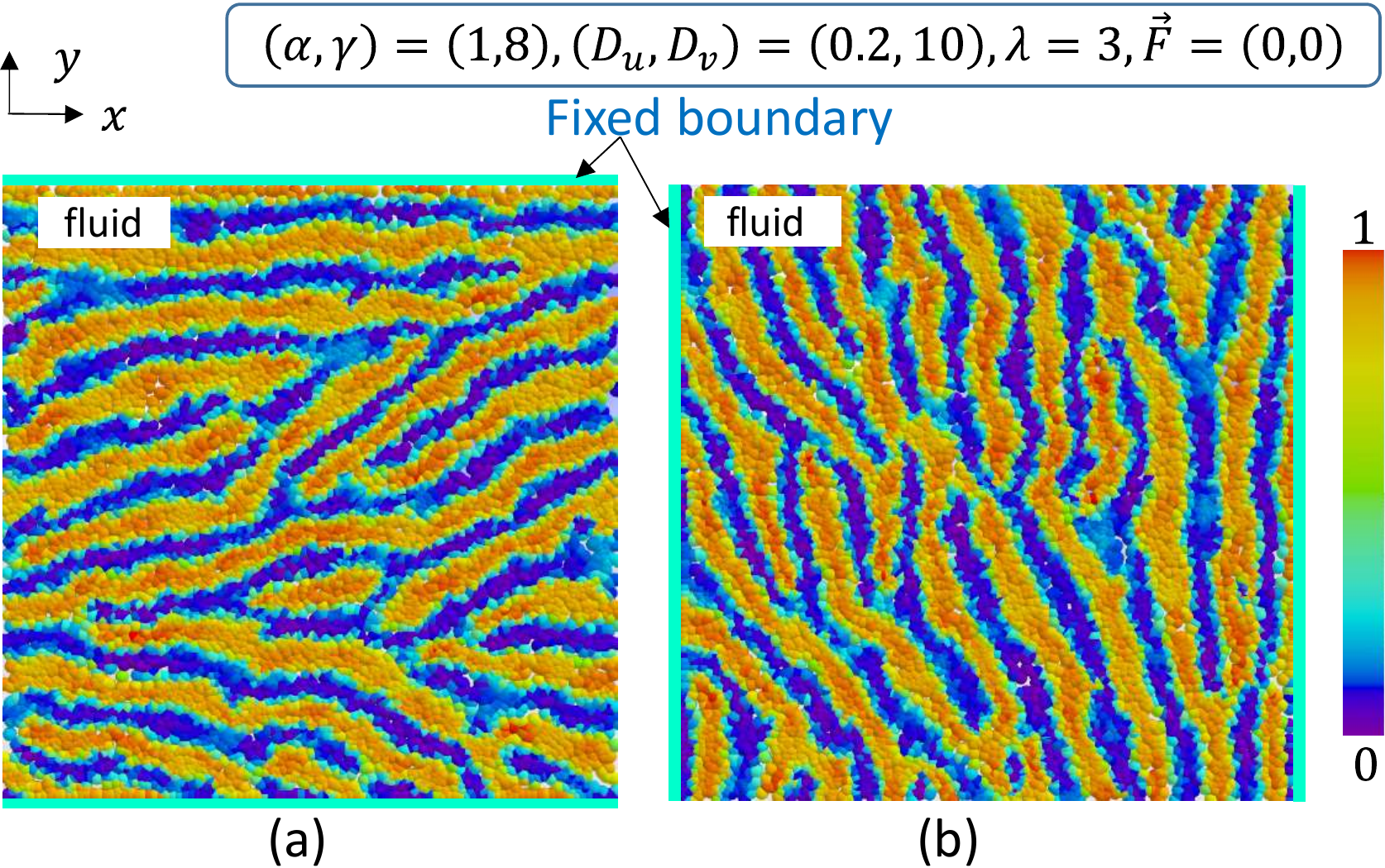}
\caption { Snapshots obtained on fluid lattices, in which a pair of fixed boundaries is assumed in (a) the $x$ direction and (b) the $y$ direction. Note that the boundary conditions assumed here and shown in Figs. \ref{fig-17}(a),(b)  are different from those used so far. The lattice size is $N\!=\!10000$, and the  parameters are shown in the figure. The alignment of TPs is a spontaneous one because of $\vec{F}\!=\!(0,0)$, and therefore, spontaneously emergent directions are determined by boundary conditions. 
\label{fig-18}}
\end{figure}
We find from Figs. \ref{fig-18}(a),(b) that TPs align along the fixed-boundary direction.

\section{Concluding remarks \label{conclusion}}
In this paper, we have successfully shown that the Finsler geometry (FG) modeling technique is applicable to a differential equation model. We numerically studied anisotropic Turing patterns (TPs) by using the FG modeling technique with an internal degree of freedom (IDOF) to implement diffusion anisotropy in the reaction-diffusion (RD) equation of FitzHugh-Nagumo (FN).
In the ordinary numerical techniques used to solve the RD equation, direction-dependent diffusion coefficients are assumed as an input to reproduce anisotropic TPs. On the other hand, anisotropy in diffusion coefficients is dynamically generated in FG modeling. In this sense, the anisotropic TPs are attributed to the IDOF alignment originating in direction-dependent fluctuations of vertices. The vertices are biologically interpreted as cells or lumps of cells, and therefore, our results indicate that one possible origin of anisotropic TPs is direction-dependent movements of cells compatible with those observed in developmental processes.

The surface tension due to the boundary frame of membranes depends on the frame area and interacts with the IDOF. For this reason, the IDOF on fixed-connectivity lattices is expected to be controlled by uniaxial strains, which can also be controlled by suitable boundary conditions. We confirm that the TP direction is controlled by the surface boundary conditions such that the direction aligns along the tensile or compressive strain direction on fixed-connectivity lattices. On fluid lattices with a pair of fixed boundaries, we confirm that anisotropic TPs spontaneously emerge along the direction of fixed boundaries. 

These anisotropic TPs are understood to be anisotropic diffusion in the FG modeling, and the diffusion rate is calculable on these fluid lattices by regarding Monte Carlo iteration as time. It is interesting to study the dependence of anisotropic TPs on the diffusion rate. This remains to be a future study.

\begin{acknowledgments}
This work is supported in part by Collaborative Research Project J20Ly19 of the Institute of Fluid Science (IFS), Tohoku University. The authors acknowledge Dr. Jean-Paul Rieu and Dr. B. Ducharne for suggestions. Numerical simulations were performed on the supercomputer system AFI-NITY at the Advanced Fluid Information Research Center, Institute of Fluid Science, Tohoku University.
\end{acknowledgments}

\appendix

\section{Computational domain spanned by a membrane with a nontrivial surface tension\label{App-A}}
In this Appendix, we give detailed information on the computational domain ${\cal D}$ bounded by a fixed frame for PBCs in Figs. \ref{fig-6}(a) and \ref{fig-8}(a). As mentioned in Section \ref{FG-modeling}, this domain ${\cal D}$ has an internal structure described by the vertex positions $\vec{r}_i (\in {\bf R}^2, 1\!\leq\!i\!\leq\!N)$, and these $\vec{r}_i$ are regarded as membranes \cite{KANTOR-NELSON-PRA1987,Gompper-Kroll-PRA1992,KOIB-PRE-2005,Ho-Baum-EPL1990,HELFRICH-1973,Peliti-Leibler-PRL1985,NELSON-SMMS2004}, which are classical mechanical $N$-particle systems governed by the spring potential or the Gaussian bond potential $S_1(\vec{r})$ and by several constraints imposed on $\vec{r}_i$. Mainly due to $S_1$ and the boundary frame, the membrane ${\cal D}$ is exposed to a tensile stress $\sigma$, which comes from the so-called scale-invariant property of the partition function, as mentioned in Section \ref{FG-modeling} \cite{Wheater-JPA1994}. This property is common to both fixed and fluid models, and hence, we use $Z_{\rm fix}$ in Eq. (\ref{part-funct}) here in this Appendix for simplicity.
\begin{figure}[h]
\centering{}\includegraphics[width=11.5cm,clip]{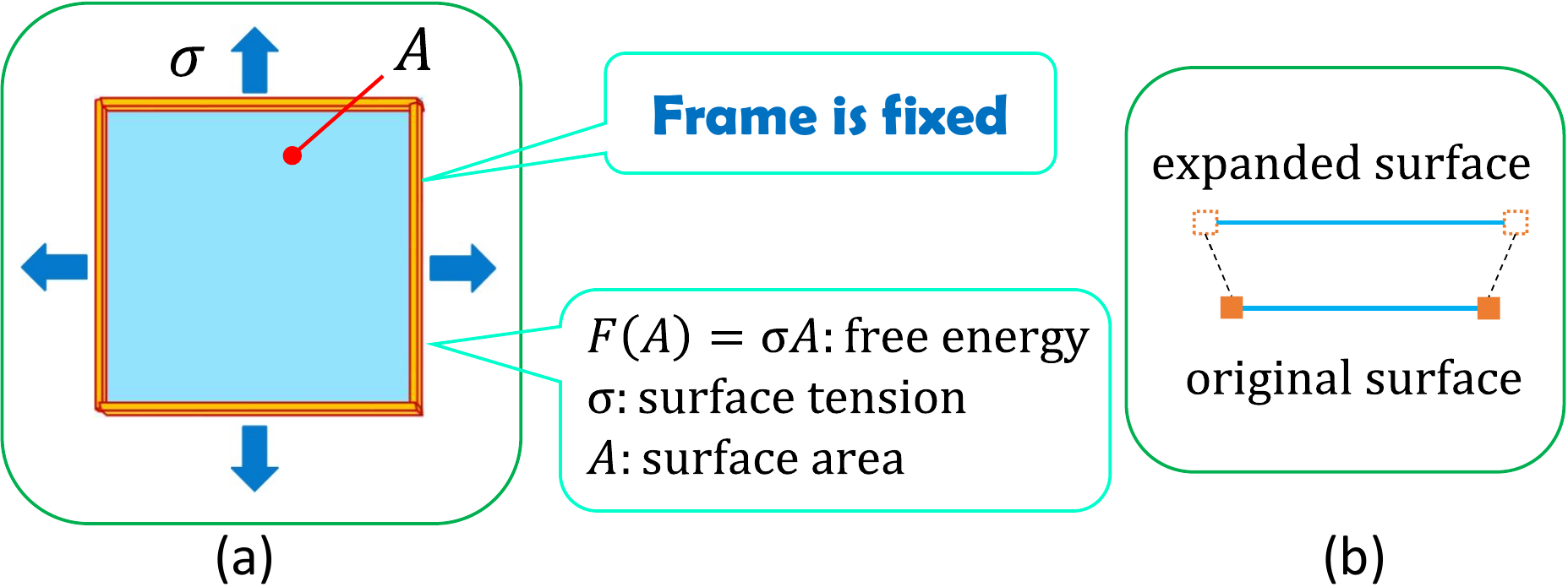}
\caption{
(a) Surface enclosed by a fixed boundary frame of area $A$ under a constant surface tension $\sigma$. This surface has a mechanical free energy $F(A)\!=\!\sigma A(=\!\int ^A \sigma dA)$ because $\sigma$ is constant and isotropic. (b) Illustrations of the original surface with a boundary frame and an expanded surface by a scale parameter $\beta(\geq\!1)$.
\label{fig-A-1} }
\end{figure}

First, we note that the variables $\vec{r}_i$ are integrated out in $Z_{\rm fix}$ such that $\int\prod_i d\vec{r}_i\exp(-S)$, corresponding to $2N$-dimensional multiple integration. Here, we rewrite $S$ as $S\!=\!S_1\!+\!U_{\rm fr}(A)\!+\!S_{\rm other}$, where $U_{\rm fr}(A)$ is the potential for fixing the frame area $A$ to $A\!=\!L_xL_y$ and not included in $S$ of Eq. (\ref{discrete-total-Hamiltonian}), as mentioned in the text, and $S_{\rm other}$ is the remaining term in $S$. Note that $S_1\!=\!\sum_{ij}\ell_{ij}^2$ and $U_{\rm fr}(A)$ depend on $\vec{r}$ and that $S_{\rm other}$ is independent of $\vec{r}$. Now, we replace $\vec{r}$ with $\vec{r}_{\beta}\!=\!\beta\vec{r}$ in $Z_{\rm fix}$ with the scale parameter $\beta(\geq 1)$. This replacement constitutes a simple variable change in the integrations $\int d\vec{r}_i \!\to\! \beta^2\int d\vec{r}_i$, $S_1\!\to\! \beta^2 S_1$, $\cdots$, and it does not change $Z_{\rm fix}$, and therefore, we have $Z_{\rm fix}(\vec{r})\!=\!Z_{\rm fix}(\vec{r}_{\beta})$, where
\begin{eqnarray}
\label{scaled-Z}
Z_{\rm fix}(\vec{r}_{\beta})=\beta^{2N}\int\prod_i d\vec{r}_i\exp\left[-\beta^{2}S_1\!-\!U_{\rm fr}(A_\beta)\!-\!S_{\rm other}\right].
\end{eqnarray}
In the expression $U_{\rm fr}(A_\beta)$, $A_\beta$ is given by $A_\beta\!=\!\beta^{-2}A$ because the fixed frame means that the area scales as $A\!\to\!A_\beta\!=\!\beta^{-2}A$ according to the scale change $\vec{r}\!\to\!\beta\vec{r}$.
By differentiating both sides of $Z_{\rm fix}(\vec{r})\!=\!Z_{\rm fix}(\vec{r}_{\beta})$ with respect to $\beta$ and by fixing $\beta\!\to\!1$, we have
\begin{eqnarray}
\label{differential-fixing}
\begin{split}
0&=\left.\frac{\partial}{\partial \beta}Z_{\rm fix}(\vec{r}_{\beta})\right|_{\beta=1}\\
&=\left.\frac{\partial}{\partial \beta}\left(\beta^{2N}\int\prod_i d\vec{r}_i\exp\left[-\beta^{2}S_1\!-\!U_{\rm fr}(A_\beta)\!-\!S_{\rm other}\right]\right)\right|_{\beta=1}\\
&=\left.2N\beta^{-1}Z_{\rm fix}(\vec{r}_{\beta})\right|_{\beta=1}+\left.2\beta^{2N+1}\int\prod_i d\vec{r}_i \,S_1\,\exp\left[-S(\vec{r}_{\beta})\right]\right|_{\beta=1}+\left.\frac{\partial Z_{\rm fix}(\vec{r}_{\beta})}{\partial A_\beta} \frac{\partial A_\beta}{\partial \beta}\right|_{\beta=1}\\
&=2NZ_{\rm fix}(\vec{r})-2\int\prod_i d\vec{r}_i \,S_1\,\exp\left[-S(\vec{r})\right]-2A\frac{\partial Z_{\rm fix}(\vec{r})}{\partial A},
\end{split}
\end{eqnarray}
where $\left.\frac{\partial A_\beta}{\partial \beta}\right|_{\beta=1}\!=\!\left.\frac{\partial \beta^{-2} A}{\partial \beta}\right|_{\beta=1}\!=\!-2A$ is used in the final term of the third line. We note that the boundary of the expanded surface by $\beta(\geq\!1)$ is identified with the fixed frame only when  $\beta\!\to\!1$ (Fig. \ref{fig-A-1}(b)).
Multiplying $2^{-1}Z_{\rm fix}(\vec{r})^{-1}$ by the final expression in Eq. (\ref{differential-fixing}), we obtain
\begin{eqnarray}
0=N-\langle S_1\rangle -\frac{A}{Z_{\rm fix}(\vec{r})}\frac{\partial Z_{\rm fix}(\vec{r})}{\partial A}.
\end{eqnarray}
Since $\langle S_1\rangle\!=\!\langle \sum_{ij}\ell_{ij}^2\rangle$ can be replaced by $\sum_{ij}\langle \ell_{ij}^2\rangle\!=\!\sum_{ij}\langle \ell^2\rangle\!=\!\langle \ell^2\rangle\sum_{ij}1\!=\!N_B\langle\ell^2\rangle\!=\!3N\langle\ell^2\rangle$, we have the mean squared bond length $\langle \ell^2\rangle $ such that
\begin{eqnarray}
 \label{mean-bond-length-squares}
\langle \ell^2\rangle =\frac{1}{3}-\frac{1}{3N}\frac{A}{Z_{\rm fix}(\vec{r})}\frac{\partial Z_{\rm fix}(\vec{r})}{\partial A}.
\end{eqnarray}
If the boundary frame for PBCs is not assumed, the second term on the right-hand side is unnecessary, and hence, we have $\langle \ell^2\rangle \!=\!{1}/{3} (\Leftrightarrow \langle S_1\rangle/N\!=\!1)$, as mentioned in the main text.

The problem is how to evaluate the second term in the case that the frame is present. The potential $U_{\rm fr}(A)$ can be expressed only as $U_V$ in Eq. (\ref{discrete-Hamiltonian}) and is therefore not differentiable.
To evaluate $\frac{\partial Z_{\rm fix}(\vec{r})}{\partial A}$ for surfaces with a fixed frame, we regard the membrane of the $N$-particle system bounded by the frame as a sheet of area $A$ without the internal structure associated with the $N$-particles. In this simple surface of area $A$, its free energy $F(A)$ is given by $F(A)\!=\!\sigma A$, where $\sigma$ is the surface tension. Therefore, its partition function can also be expressed as $Z(A)\!=\!\exp[-F(A)]$. Thus, ${\partial Z_{\rm fix}(\vec{r})}/{\partial A}$ can be evaluated by replacing $Z_{\rm fix}(\vec{r})$ with $\exp[-F(A)]\!=\!\exp(-\sigma A)$, and we have $\langle \ell^2\rangle \!=\!1/3\!+\!\sigma A/(3N)$ from Eq. (\ref{mean-bond-length-squares}), and therefore
\begin{eqnarray}
 \label{surface-tension}
\sigma=\frac{3N}{A}\left(\langle\ell^2\rangle -\frac{1}{3}\right)\simeq\frac{3}{d^2}\left(\langle\ell^2\rangle -\frac{1}{3}\right), \quad A=L_xL_y=(n_x-1)(n_y-1)d^2,
\end{eqnarray}
where $L_x, L_y$ and $n_x, n_y$ are shown in Fig. \ref{fig-6}(a). Thus, $\sigma$ depends on the lattice spacing $d$.

We note that the scale-dependent constraint $U_V$ in Eq. (\ref{discrete-Hamiltonian}) should be included in the $\exp$ term of Eq. (\ref{scaled-Z}) such that $\exp\left[-\beta^{2}S_1\!-\!U_{\rm fr}(A_\beta)\!-\!U_V(\ell_{\rm min}^\beta, \ell_{\rm max}^\beta)\!-\!S_{\rm other}\right]$ with $\ell_{\rm min}^\beta\!=\!\beta^{-1}\ell_{\rm min}$ and $\ell_{\rm max}^\beta\!=\!\beta^{-1}\ell_{\rm max}$. Correspondingly, the terms $-\ell_{\rm min}\frac{\partial Z_{\rm fix}(\vec{r})}{\partial \ell_{\rm min}}-\ell_{\rm max}\frac{\partial Z_{\rm fix}(\vec{r})}{\partial \ell_{\rm max}}$ are included on the right-hand side of Eq. (\ref{differential-fixing}). However, these terms can be dropped when $Z_{\rm fix}(\vec{r})$ is identified with $Z(A)\!=\!\exp(-\sigma A)$, which has no internal structure and is independent of $\ell_{\rm max}$ and $\ell_{\rm min}$. For this reason, we exclude $U_V$ from the $\exp$ of Eq. (\ref{scaled-Z}) from the beginning.

\section{Finsler function and Finsler length \label{App-B}}
\begin{figure}[h]
\centering{}\includegraphics[width=11.5cm,clip]{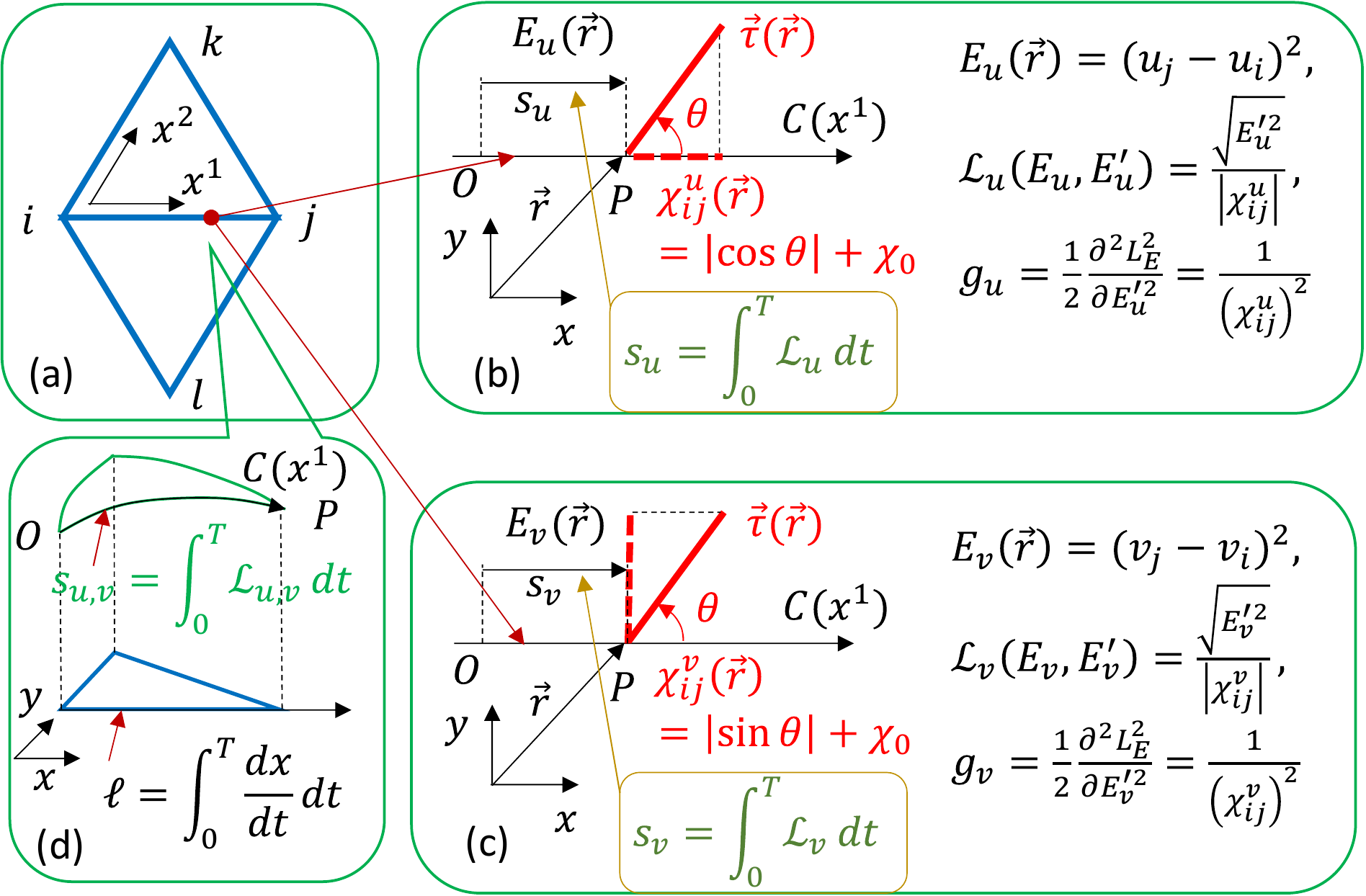}
\caption{
(a) A local coordinate system $(x^1,x^2)$ with the origin at vertex $i$ on a triangle $\bigtriangleup_{ijk}$, (b) unit Finsler length $v^{u}_{ij}$ for the diffusion anisotropy of $u$ on the curve $C(x^1)$ along the $x^1$-axis, (c) unit Finsler length $v^{v}_{ij}$ corresponding to the variable $v$, and (d) intuitive illustrations of Euclidean length $\ell$ and Finsler length $s_{u,v}$ between $OP$ along the $x^1$- and $C(x^1)$-axes, respectively. The direction-dependent Finsler metrics $g_u$ and $g_v$ along the $x^1$-axis, which is assumed to be parallel to the $x$-axis, from $i$ to $j$ are obtained by using Finsler functions ${\cal L}_u$ and ${\cal L}_v$ such that $g_u\!=\!1/(\chi_{ij}^u)^{2}$ and $g_v\!=\!1/(\chi_{ij}^v)^{2}$. Arrows with $x$ and $y$ in (b), (c) and (d) represent the canonical coordinate axes of ${\bf R}^2$.
\label{fig-A-2} }
\end{figure}
In this Appendix, we introduce technical details of Finsler metrics for diffusion anisotropy. The Finsler length can be introduced in a local coordinate system $(x^1,x^2)$ on triangulated lattices in a discrete manner (Fig. \ref{fig-A-2}(a)). An origin of the local coordinate is the vertex position $i$ of the triangle $\bigtriangleup_{ijk}$. The $x^1$-axis is considered a curve $C(x^1)$, on which a Finsler function ${\cal L}$ is introduced using a coordinate function $E_u(\vec{\bf r})\!=\!E_u(x^1)$ and its first derivative $E_u^\prime(x^1)\!=\!\frac{dE_u}{dx^1}$ such that
\begin{eqnarray}
\label{Finsler-function-phi}
\begin{split}
 &{\cal L}_u\left(E_u,E_u^\prime\right)=\frac{\sqrt{E_u^{\prime\;2}}}{|\chi_u|}, \\
 & E_u=\int_0^{x^1} \left(\frac{\partial u}{\partial x^1} \right)^2dx^1, \quad
    E_u^\prime=\frac{dE_u}{dx^1}.
\end{split}
\end{eqnarray}
Here, we use the expression $\left(\frac{\partial u}{\partial x^1} \right)^2\!=\!(\nabla u)^2$ along $C(x^1)$. $E_u(x^1)$ is an energy function that is positive and increases with increasing $(\nabla u)^2$ along the $x^1$-axis and monotonically increases with respect to $x^1$; hence, $E_u(x^1)$ can be used to define the Finsler length along $C(x^1)$. Note that this coordinate function $E_u(\vec{\bf r})\!=\!E_u(x^1)$ is different from those in Ref. \cite{Krause-etal-BulMatBiol2019}, where the induced metric is used, and hence, the coordinate functions are defined on the tangential lines of Euclidean length on curved surfaces.
From the general prescription, we obtain the Finsler length $s$ by the time integration of the Finsler function ${\cal L}$ along $C(x^1)$ from $t\!=\!0$ to $t\!=\!T$ such that $s\!=\!\int _0^T {\cal L}(E_u,E_u^\prime)dt$. Equivalently, we obtain the corresponding Finsler metric element $g^u$ along the $x^1$-axis such that $g^u\!=\!(1/2)\partial ^2{\cal L}_u^2/\partial E_u^{\prime\;2}\!=\!1/(\chi_{ij}^u)^{2}$, which is suitable for our purpose. Another Finsler metric for anisotropic diffusion of the variable $v$ along $C(x^1)$ can also be obtained by $g^v\!=\!(1/2)\partial ^2{\cal L}_v^2/\partial E_v^{\prime\;2}\!=\!1/(\chi_{ij}^v)^{2}$. Here, we assume that $\chi_{ij}^u$ and $\chi_{ij}^v$ are given by
\begin{eqnarray}
\label{Finsler-unit-lengths-App}
\begin{split}
&\chi_{ij}^u=|\vec{\tau}_i\cdot\vec{e}_{ij}| + \chi_0,\\
&\chi_{ij}^v=\sqrt{1-|\vec{\tau}_i\cdot\vec{e}_{ij}|^2} + \chi_0,
\end{split}
\end{eqnarray}
where $\vec{\tau}_i$ denotes the internal DOF at vertex $i$, $\vec{e}_{ij}$ is the unit tangential vector from vertices $i$ to $j$, and $\chi_0$ is a small positive number. These parameters $\chi_{ij}^u$ and $\chi_{ij}^v$ on $C(x^1)$ are visualized by the dashed lines in Fig. \ref{fig-A-2}(b) and (c).

Here, we note that the relation between the Euclidean length $dx^1$ between $i$ and $j$ and the Finsler length $ds_u$ from $i$ to $j$ is given by using the Finsler metric $g_u$ such that $(ds_u)^2\!=\!g_u(dx^1)^2\!=\!(dx^1)^2/(\chi_{ij}^u)^{2}$. Thus, we have $ds_u\!=\!dx^1/\chi_{ij}^u$, and therefore $ds_u\!=\!1$ for $dx^1\!=\!\chi_{ij}^u$. This is the reason we call $\chi_{ij}^u$ the unit Finsler length from $i$ to $j$. We sometimes call $\chi_{ij}^u$ a ``velocity,'' as $\chi_{ij}^u$ physically has the units of velocity because the Finsler length $ds_u$ is a time length.
Along the $x^2$-axis from vertices $i$ to $k$ in Fig. \ref{fig-A-2}(a), the unit Finsler lengths $\chi_{ik}^u$ and $\chi_{ik}^v$ can be defined by replacing $\vec{e}_{ij}$ with $\vec{e}_{ik}$ in Eq. (\ref{Finsler-unit-lengths-App}). Thus, we obtain the two-dimensional Finsler metrics $g^u$ and $g^v$ with respect to the local coordinate $(x^1,x^2)$ such that
\begin{eqnarray}
\label{Finsler-metric-App}
g^u= 
\begin{pmatrix}
1/(\chi_{ij}^u)^{2} & 0 \\ 
0 & 1/(\chi_{ik}^u)^{2} 
\end{pmatrix}, 
  \quad
 g^v= 
\begin{pmatrix}
1/(\chi_{ij}^v)^{2} & 0  \\
0 & 1/(\chi_{ik}^v)^{2} 
\end{pmatrix}.
\end{eqnarray}

\section{Discrete Hamiltonian \label{App-C}}
\begin{figure}[h]
\centering{}\includegraphics[width=11.5cm,clip]{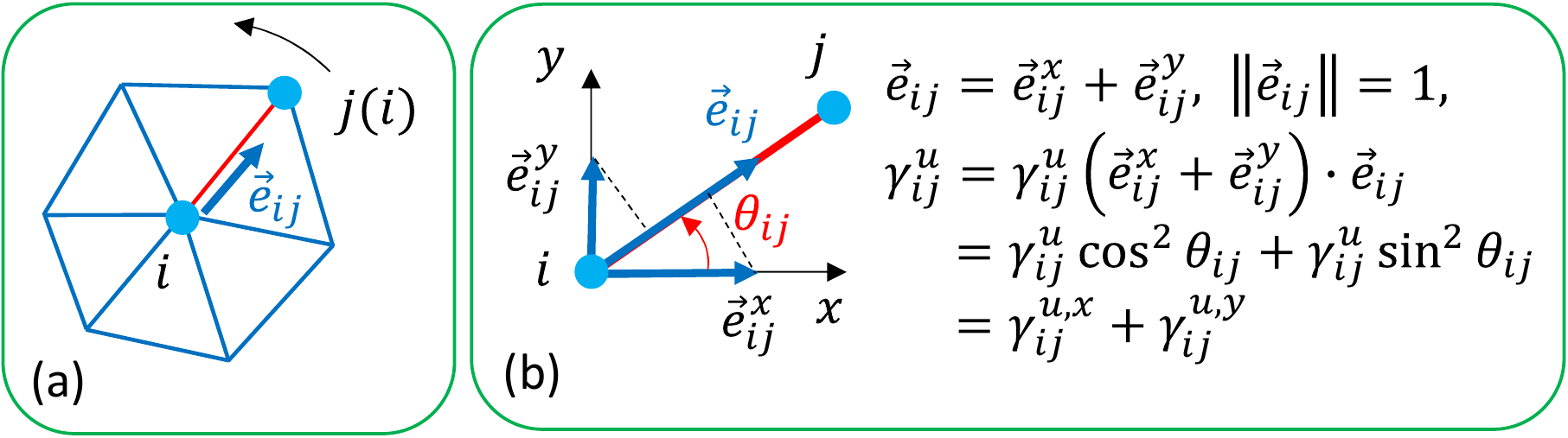}
\caption{
(a) Vertices $j(i)$ connected to vertex $i$ and (b) unit tangential vector $\vec{e}_{ij}$ from vertices $i$ to $j(i)$ and its decomposition $\vec{e}_{ij}\!=\!\vec{e}_{ij}^{\,x}\!+\!\vec{e}_{ij}^{\,y}$ into the canonical coordinate axis components.
\label{fig-A-3} }
\end{figure}
Using $g_u$ and $g_v$ in Eq. (\ref{Finsler-metric-App}), we have the discrete expression of $S_u$ and $S_v$ in Eq. (\ref{discrete-Hamiltonian}). Here, we show the outline of discretization of $S_u$ as follows: The integral is replaced by the sum over triangles with the determinant $g\!=\!\det(g_u)$ such that $\int \sqrt{g_u}d^2x \to \sum_{\bigtriangleup} 1/(\chi_{ij}^u \chi_{ik}^u)$, and the differentials are replaced by differences with the inverse metric $g_u^{ab}\!=\!(g_u)^{-1}$ such that $g_u^{ab}\frac{\partial u}{\partial x^a}\frac{\partial u}{\partial x^b}\!=\!g_u^{11}(\frac{\partial u}{\partial x^1})^2\!+\!g_u^{22}(\frac{\partial u}{\partial x^2})^2$ $\to$ $(\chi_{ij}^u)^2(u_j-u_i)^2\!+\!(\chi_{ik}^u)^2(u_k-u_i)^2$. Thus, we have $S_u\!=\!\frac{1}{2}\int \sqrt{g_u}d^2x g_u^{ab}\frac{\partial u}{\partial x^a}\frac{\partial u}{\partial x^b}$ $\to$ $\frac{1}{2}\sum_{\bigtriangleup} \left(\frac{\chi_{ij}^u}{\chi_{ik}^u}(u_j-u_i)^2\!+\!\frac{\chi_{ik}^u}{\chi_{ij}^u}(u_k-u_i)^2 \right)$. Here, we note that there are two different origins of local coordinates at vertices $j$ and $k$ other than $i$ on triangle $\bigtriangleup_{ijk}$. The discrete expressions of $S_u$ corresponding to these local coordinates on $\bigtriangleup_{ijk}$ are obtained by replacing the indices $ijk \to jki$ and $jki \to kij$ and summing over three different expressions with the factor $1/3$, we obtain
\begin{eqnarray}
\label{discrete-Sphi-1}
\begin{split}
S_u&=\frac{1}{6}\sum_{\bigtriangleup_{ijk}} \left[\frac{\chi_{ij}^u}{\chi_{ik}^u}(u_j-u_i)^2+\frac{\chi_{ik}^u}{\chi_{ij}^u}(u_k-u_i)^2 +\frac{\chi_{jk}^u}{\chi_{ji}^u}(u_k-u_j)^2 \right.\\
&\left. +\frac{\chi_{ji}^u}{\chi_{jk}^u}(u_i-u_j)^2+\frac{\chi_{ki}^u}{\chi_{kj}^u}(u_i-u_k)^2+\frac{\chi_{kj}^u}{\chi_{ki}^u}(u_j-u_k)^2\right]\\
&=\sum_{\bigtriangleup_{ijk}} \left(\bar{\gamma}_{ij}^u(u_i-u_j)^2+\bar{\gamma}_{jk}^u(u_j-u_k)^2 +\bar{\gamma}_{ki}^u(u_k-u_i)^2\right), \\
&\bar{\gamma}_{ij}^u=\frac{1}{6}\left(\frac{\chi_{ij}^u}{\chi_{ik}^u}+\frac{\chi_{ji}^u}{\chi_{jk}^u}\right),\quad \bar{\gamma}_{jk}^u=\frac{1}{6}\left(\frac{\chi_{jk}^u}{\chi_{ji}^u}+ \frac{\chi_{kj}^u}{\chi_{ki}^u}\right), \quad
\bar{\gamma}_{ki}^u=\frac{1}{6}\left(\frac{\chi_{ki}^u}{\chi_{kj}^u}+\frac{\chi_{ik}^u}{\chi_{ij}^u}\right),
\end{split}
\end{eqnarray}
where the sum over triangles is explicitly denoted by $\sum_{\bigtriangleup_{ijk}}$.
The summation convention can also be changed from summation over triangles to summation over bonds $\sum_{ij}$. Recalling that the bond $ij$ is shared by two triangles $\bigtriangleup_{ijk}$ and $\bigtriangleup_{jil}$ (Fig. \ref{fig-A-2}(a)), we have to include terms $\frac{1}{6}\left(\frac{\chi_{ij}^u}{\chi_{ik}^u}+\frac{\chi_{ji}^u}{\chi_{jk}^u}\right)(u_i-u_j)^2$ from $\bigtriangleup_{ijk}$ and $\frac{1}{6}\left(\frac{\chi_{ji}^u}{\chi_{jl}^u}+\frac{\chi_{ij}^u}{\chi_{il}^u}\right)(u_j-u_i)^2$ from $\bigtriangleup_{jil}$ in the sum over bonds $\sum_{ij}$. Thus, we have
\begin{eqnarray}
\label{discrete-Sphi-2}
\begin{split}
S_u=\sum_{ij} \gamma_{ij}^u(u_i-u_j)^2, \quad
\gamma_{ij}^u=\frac{1}{6}\left(\frac{\chi_{ij}^u}{\chi_{ik}^u}+\frac{\chi_{ji}^u}{\chi_{jk}^u}+\frac{\chi_{ij}^u}{\chi_{il}^u}+\frac{\chi_{ji}^u}{\chi_{jl}^u}\right), \quad \gamma_{ij}^{u}=\gamma_{ji}^{u}.
\end{split}
\end{eqnarray}

Now, we describe the outline of the discretization of the Laplace Beltrami operator.
\begin{eqnarray}
\label{Laplace-Beltrami}
 \Laplace =\frac{1}{\sqrt{g}}\frac{\partial}{\partial x_a}\left(\sqrt{g}g^{ab}\frac{\partial  }{\partial x^b}\right),
\end{eqnarray}
on triangulated lattices. Since this operator includes second-order differentials, we adopt an indirect discretization scheme based on the discrete Hamiltonian $S_u$ in Eq. (\ref{discrete-Sphi-1}). The continuous expression of $\Laplace u$ can be obtained by the variational technique for the continuous $S_u$:
\begin{eqnarray}
\label{variation-continuous}
S_u\!=\!\frac{1}{2}\int \sqrt{g_u}d^2x g_u^{ab}\frac{\partial u}{\partial x^a}\frac{\partial u}{\partial x^b} \quad \xrightarrow{\rm{variation}}\quad \Laplace u =-\frac{\partial S_u}{\partial u}=\frac{1}{\sqrt{g}}\frac{\partial}{\partial x_a}\left(\sqrt{g}g^{ab}\frac{\partial  u}{\partial x^b}\right).
\end{eqnarray}
Therefore, we obtain the discrete expression $\Laplace u_i$ from $S_u$ in Eq. (\ref{discrete-Sphi-1}) by the discrete variational technique
\begin{eqnarray}
\label{variation-discrete}
S_u=\sum_{ij} \gamma_{ij}^u(u_i-u_j)^2 \quad \xrightarrow{\rm{variation}}\quad \Laplace u_i =-\frac{\partial S_u}{\partial u_i} =-\frac{\partial}{\partial u_i}  \sum_{jk} \gamma_{jk}^u(u_j-u_k)^2
\end{eqnarray}
such that
\begin{eqnarray}
\label{discrete-Laplace-App}
\Laplace u_i = 2\left(\sum_{j(i)} \gamma_{ij}^u u_j -u_i\sum_{j(i)} \gamma_{ij}^u\right)=\sum_{j(i)}2\gamma_{ij}^u\left(u_j-u_i \right),
\end{eqnarray}
where $\sum_{j(i)} $ denotes the sum over vertices $j$ connected to vertex $i$ with bond $ij$ (Figs. \ref{fig-A-3}(a) and \ref{fig-6}(b)). To perform this summation, we replace the sum $\sum_{jk}$ over bonds $jk$ in Eq. (\ref{variation-discrete}) with $\frac{1}{2} \sum_{j}\sum_{k(j)}(=\!\sum_{jk})$, where $\sum_{j}$ and $\sum_{k(j)}$ are the sum over vertices $j$ and the sum over vertices $k$, which is connected to $j$ with bond $jk$, respectively. The factor $1/2$ appears due to the duplicated sums in $\sum_{j}\sum_{k(j)}$; the term with index $jk\!=\!12(\Leftrightarrow j\!=\!1,k\!=\!2)$ appears twice in $\sum_{j}\sum_{k(j)}$, for example. Note that $\sum_{j}\sum_{k(j)}\!=\!\sum_{k}\sum_{j(k)}$ because $\sum_{jk}\!=\!\sum_{kj}$. Thus, we have
\begin{eqnarray}
\label{discretization-Laplace}
\begin{split}
&\frac{\partial}{\partial u_i}  \sum_{jk} \gamma_{jk}^u(u_j-u_k)^2 
=\frac{\partial}{\partial u_i}  \sum_{jk} \gamma_{jk}^u(u_j^2+u_k^2-2u_ju_k)\\
&= \frac{1}{2}\sum_{j} \sum_{k(j)}\gamma_{jk}^u 2(u_j\delta_{ji}+u_k\delta_{ki}-\delta_{ij}u_k-u_j\delta_{ik}) \\
&=\sum_{j} \sum_{k(j)}\gamma_{jk}^uu_j\delta_{ji}+\sum_{j} \sum_{k(j)}\gamma_{jk}^uu_k\delta_{ki}-\sum_{j} \sum_{k(j)}\gamma_{jk}^u\delta_{ij}u_k-\sum_{j} \sum_{k(j)}\gamma_{jk}^uu_j\delta_{ik},
\end{split}
\end{eqnarray}
where $\delta_{ij}$ is the Kronecker delta. In this final expression, the first term is $\sum_{j} \sum_{k(j)}\gamma_{jk}^uu_j\delta_{ji}\!=\!\sum_{k(i)}\gamma_{ik}^uu_i\!=\!u_i\sum_{k(i)}\gamma_{ik}^u$, and the second term can be written as
$\sum_{j} \sum_{k(j)}\gamma_{jk}^uu_k\delta_{ki}\!=\!\sum_{k} \sum_{j(k)}\gamma_{jk}^uu_k\delta_{ki}\!=\!\sum_{j(i)}\gamma_{ji}^uu_i\!=\!u_i\sum_{j(i)}\gamma_{ji}^u$, which is identical to the first term. The third term is $-\!\sum_{j} \sum_{k(j)}\gamma_{jk}^u\delta_{ij}u_k\!=\!-\!\sum_{k(i)}\gamma_{ik}^uu_k$, which is identical to the fourth term $-\!\sum_{j} \sum_{k(j)}\gamma_{jk}^uu_j\delta_{ik}\!=\!-\!\sum_{k} \sum_{j(k)}\gamma_{jk}^uu_j\delta_{ik}\!=\!-\!\sum_{j(i)} \gamma_{ji}^uu_j\!=\!-\!\sum_{j(i)} \gamma_{ij}^uu_j$. This proves that Eq. (\ref{discrete-Laplace-App}).

Here, we comment on the relation between $\Laplace$ in Eq. (\ref{discrete-Laplace-App}) and the network Laplacian $\sum_{j=1}^N L_{ij}u_j\!=\!\sum_{j=1}^N A_{ij}(u_j\!-\!u_i)$, where $N$ is the total number of nodes (or vertices) and
$L_{ij}\!=\!A_{ij}-k_i\delta_{ij}$ with $k_i\!=\!\sum_{j=1}^NA_{ij}$. The adjacency (or connectivity) matrix $A_{ij}$ is defined by $A_{ij}\!=\!1$ if $i$ and $j$ are connected and $A_{ij}\!=\!0$ otherwise. Rewriting  $\Laplace u_i$ in Eq.(\ref{discrete-Laplace-App}) as
\begin{eqnarray}
\label{network-Laplacian}
\begin{split}
\Laplace u_i=\sum_{ij}2\gamma_{ij}^u \left(A_{ij}-k_i\delta_{ij}\right),
\end{split}
\end{eqnarray}
we find that $\Laplace u_i$ is considered to be a weighted network Laplacian with the weight $2\gamma_{ij}^u$. Note that $\gamma_{ij}^u$ is well defined for networks on two-dimensional surfaces, as shown in Fig. \ref{fig-8}.
In the isotropic case, $\chi_0\to\infty$ in Eq. (\ref{Finsler-unit-lengths-App}), we have $\chi_{ij}^u/\chi_{kl}^u\!\to\! 1$ and $\gamma_{ij}\!\to\!2/3$ from Eq. (\ref{discrete-Sphi-2}), and therefore, $2\gamma_{ij}^u\!\to\! 4/3$. Thus, with a suitable normalization factor in the definition of $\gamma_{ij}^u$ in Eq. (\ref{discrete-Sphi-2}), $\Laplace u_i$ in Eq. (\ref{network-Laplacian}) reduces to the standard network Laplacian $\Laplace u_i\to\sum_{ij}\left(A_{ij}-k_i\delta_{ij}\right)$.

\section{Direction-dependent diffusion constants and effective density energies\label{App-D}}
In this Appendix, we describe direction-dependent diffusion constants and corresponding effective direction-dependent energies obtained by the discrete Laplace operator in Eq. (\ref{discrete-Laplace}). We simply call these energies ``density energies'' because the variables $u$ and $v$ represent the density of the activator and inhibitor, respectively. First, we note that the unit tangential vector $\vec{e}_{ij}$ from vertices $i$ to $j$ can be decomposed into $\vec{e}_{ij}\!=\!\vec{e}_{ij}^{\,x}\!+\!\vec{e}_{ij}^{\,y}$ (Fig. \ref{fig-A-3}(b)). Then, by using the angle $\theta_{ij}$ and the relations $1\!=\!\vec{e}_{ij}\cdot\vec{e}_{ij}\!=\!(e_{ij}^x)^2\!+\!(e_{ij}^y)^2\!=\!\cos^2\theta_{ij}\!+\!\sin^2\theta_{ij}$ satisfying $\cos^2\theta_{ij}\!=\!\sin^2\theta_{ij}\!=\!1/2$ for $\theta_{ij}\!=\!\pi/4$, it is easy to see that $\gamma_{ij}^{u}$ is decomposed into two different parts, which can be considered the $x$ and $y$ components such that
\begin{eqnarray}
\label{gamma-decomposition}
\begin{split}
 \gamma_{ij}^u=\gamma_{ij}^{u,x}+\gamma_{ij}^{u,y}=\gamma_{ij}^{u}\cos^2\theta_{ij}+\gamma_{ij}^{u}\sin^2\theta_{ij}.
\end{split}
\end{eqnarray}
This $\gamma_{ij}^u$ is considered to have values on bond $ij$. Here, we assume an isotropic case in which $\gamma_{ij}^u$ is very large on bonds that are almost parallel to the $x$-axis ($\Leftrightarrow \theta_{ij}\!\simeq\!0$ in Fig. \ref{fig-A-3}(b)), and additionally, $\gamma_{ij}^u$ is assumed to be very small on bonds that are almost parallel to the $y$-axis ($\Leftrightarrow \theta_{ij}\!\simeq\!\pi/2$ in Fig. \ref{fig-A-3}(b)). In this case, it is easy to understand that $\gamma_{ij}^{u,x}\!>\!\gamma_{ij}^{u,y}$ on bonds $ij$ almost parallel to the $x$-axis because $\cos^2\theta_{ij}\!\simeq\!1$ and $\sin^2\theta_{ij}\!\simeq\!0$, and that $\gamma_{ij}^{u,x}\!<\!\gamma_{ij}^{u,y}$ on bonds $ij$ almost parallel to the $y$-axis because $\cos^2\theta_{ij}\!\simeq\!0$ and $\sin^2\theta_{ij}\!\simeq\!1$. Therefore, such a direction dependence of $\gamma_{ij}^u$ is considered to be reflected in $\gamma_{ij}^{u,x}$ and $\gamma_{ij}^{u,y}$. Moreover, the mean values of $\gamma_{ij}^{u,x}$ and $\gamma_{ij}^{u,y}$, denoted by $\langle \gamma_{ij}^{u,x}\rangle$ and $\langle\gamma_{ij}^{u,y}\rangle$, become isotropic in the case that $\gamma_{ij}^{u}$ is constant or randomly distributed over the lattice. Indeed, in the case of constant $\gamma_{ij}^{u}$, the lattice averages $\langle \gamma_{ij}^{u,x}\rangle$ and $\langle\gamma_{ij}^{u,y}\rangle$ depend only on the mean values $\langle\cos^2\theta_{ij}\rangle$ and $\langle\sin^2\theta_{ij}\rangle$, which are the same on the regular square lattice and on the regular hexagonal lattice. On randomly triangulated lattices, $\langle\cos^2\theta_{ij}\rangle\!\simeq\!\langle\sin^2\theta_{ij}\rangle$. In the case of randomly distributed $\gamma_{ij}^{u}$, it is also natural to expect that $\langle\gamma_{ij}^{u}\cos^2\theta_{ij}\rangle\!\simeq\!\langle\gamma_{ij}^{u}\sin^2\theta_{ij}\rangle$ on both regular-triangular and random lattices. Thus, we have direction-dependent diffusion constants defined by
\begin{eqnarray}
\label{effective-diffusion-constants-1}
D_{u}^x=\frac{1}{N_B}\sum_{ij}\gamma_{ij}^{u}\cos^2\theta_{ij}=\langle\gamma_{ij}^u\cos^2\theta_{ij}\rangle, \quad
D_{u}^y=\frac{1}{N_B}\sum_{ij}\gamma_{ij}^{u}\sin^2\theta_{ij}=\langle\gamma_{ij}^u\sin^2\theta_{ij}\rangle, 
\end{eqnarray}
where $N_B\!=\!\sum_{ij}1(=\!3N)$ denotes the total number of bonds.

It is also possible to define effective density energies, corresponding to $D_u^x$ and $D_u^y$ in Eq. (\ref{effective-diffusion-constants-1}), such that
\begin{eqnarray}
\label{effective-energy-Sphi-1}
\begin{split}
&S_u^x=\frac{1}{\langle\gamma_{ij}^u\cos^2\theta_{ij}\rangle}\sum_{ij} \gamma_{ij}^u\cos^2\theta_{ij}(u_i-u_j)^2, \\
&S_u^y=\frac{1}{\langle\gamma_{ij}^u\sin^2\theta_{ij}\rangle}\sum_{ij} \gamma_{ij}^u\sin^2\theta_{ij}(u_i-u_j)^2.
\end{split}
\end{eqnarray}
Indeed, $S_u$ in Eq. (\ref{discrete-Sphi-2}) can be expressed as
\begin{eqnarray}
\label{effective-energy-Sphi-2}
\begin{split}
S_u&=\sum_{ij} \gamma_{ij}^u(u_i-u_j)^2=\sum_{ij} \left( \gamma_{ij}^u\cos^2\theta_{ij}+ \gamma_{ij}^u\sin^2\theta_{ij}\right)(u_i-u_j)^2\\
&=\langle\gamma_{ij}^u\cos^2\theta_{ij}\rangle\frac{\sum_{ij} \gamma_{ij}^u\cos^2\theta_{ij}(u_i-u_j)^2}{\langle\gamma_{ij}^u\cos^2\theta_{ij}\rangle}+\langle\gamma_{ij}^u\sin^2\theta_{ij}|\rangle\frac{\sum_{ij} \gamma_{ij}^u\sin^2\theta_{ij}(u_i-u_j)^2}{\langle\gamma_{ij}^u\sin^2\theta_{ij}|\rangle}\\
&=D_{u}^x S_u^x+D_{u}^y S_u^y.
\end{split}
\end{eqnarray}

\section{Bond-length distribution \label{App-E}}
\begin{figure}[h!]
\centering{}\includegraphics[width=10.5cm]{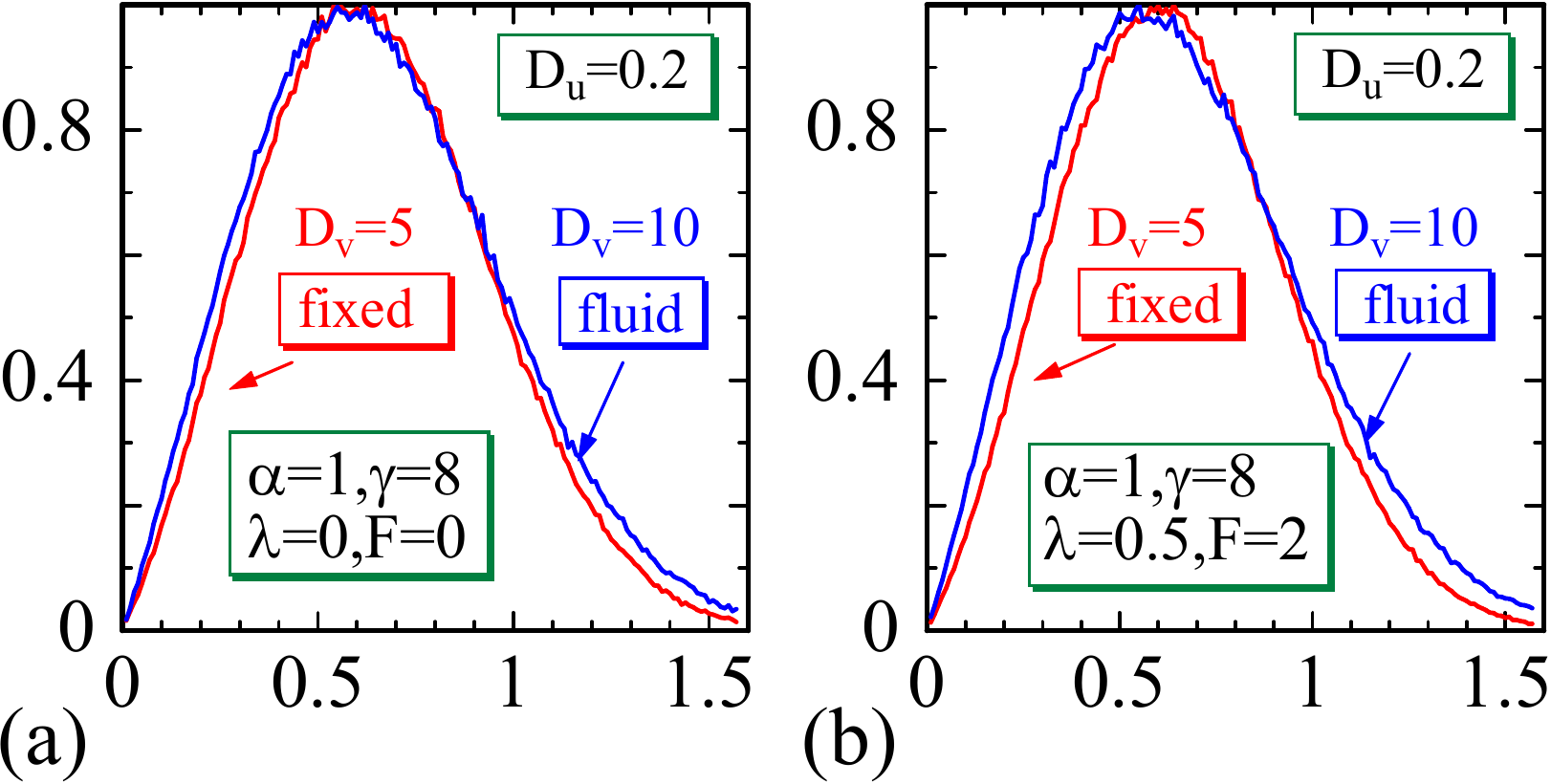}
\caption {Bond-length distribution  $h(\ell)$ vs. $\ell$ of the fixed model (red line) and fluid model (blue line) with (a) $\lambda\!=\!0$, $\vec{F}\!=\!(0,0)$ and (b) $\lambda\!=\!0.5$, $\vec{F}\!=\!(0,2)$, which respectively correspond to the isotropic and anisotropic patterns plotted in Figs. \ref{fig-9}(a), \ref{fig-10}(a) and Figs. \ref{fig-9}(d), \ref{fig-10}(d). The diffusion coefficients are $(D_u,D_v)\!=\!(0.2,5)$ for the fixed model and $(D_u,D_v)\!=\!(0.2,10)$ for the fluid model, and the other parameters $(\alpha,\gamma)\!=\!(1,8)$ are the same for the two different models. The cutoff is $\ell_{\rm max}\!=\!3d\!=\!1.575$, where $d(=\!0.525)$ is the assumed lattice spacing. The curves $h(\ell)$ vs. $\ell$ for $d\!=\!0.41$ are almost the same as those in the figures.
\label{fig-A-4}}
\end{figure}
In this Appendix, we plot the bond length distribution $h(\ell)$ on the fixed and fluid lattices of size $N\!=\!6400$ in Fig. \ref{fig-A-4}(a), (b). The lattice size is the same as that assumed for the calculations of the diffusion constants $D_u^\mu$, $D_v^\mu$ and $\sigma$ as plotted in Figs. \ref{fig-11}, \ref{fig-12} and \ref{fig-13}, respectively. The total numbers of iterations $n_{\rm itr}$ and $n_{\rm MC}$ are also the same as those in Eq. (\ref{nitr-and-N}). The coefficients are $(D_u,D_v)\!=\!(0.2,5)$ on the fixed model, while $(D_u,D_v)\!=\!(0.2,10)$ on the fluid model. These parameters, including $(\alpha,\gamma)$, are the same as those assumed in the calculation of isotropic and anisotropic snapshots in Fig. \ref{fig-9}(a) and (d) for the fixed model and in Fig. \ref{fig-10}(a) and (d) for the fluid model. We find from both isotropic and anisotropic cases in (a) and (b) that the distribution of $h(\ell)$ on the fluid lattice is slightly wide in both directions of small and large $\ell$. This slightly broad spectrum of $h(\ell)$ in the fluid model is due to the free diffusion of vertices, implying that the external force $\vec{F}$ influences the movement of vertices more strongly on the fluid lattice than on the fixed lattice. In other words, the vertex movement, which involves free diffusion, on fluid lattices is caused by a relatively weak external force $\vec{F}$.

\nocite{*}
\bibliography{aipsamp}

\end{document}